\documentclass[11pt]{article} 
\usepackage{graphicx} 
 
\newcommand{\be}{\begin{equation}} 
\newcommand{\ee}{\end{equation}} 
\newcommand{\bea}{\begin{eqnarray}} 
\newcommand{\eea}{\end{eqnarray}}

\newcommand{\parp}{\partial}

\title{Universality in the Screening Cloud of Dislocations Surrounding a Disclination}

\author{Alex Travesset \\
Physics Department\\ 
Iowa State University and Ames National Lab \\
Ames, IA, 50011-3160, USA \\  }

\begin{document} 

\date{}

\maketitle

\begin{abstract}

A detailed analytical and numerical analysis for the dislocation cloud surrounding
a disclination is presented. The analytical results show that the combined system behaves
as a single disclination with an effective fractional charge which can be 
computed from the properties
of the grain boundaries forming the dislocation cloud. Expressions are also given when
the crystal is subjected to an external two-dimensional pressure. The analytical 
results are generalized to a scaling form for the energy which up to core energies is given 
by the Young modulus of the crystal times a universal function. The accuracy of 
the universality hypothesis is numerically checked to high accuracy. The numerical approach, 
based on a generalization from previous work by S. Seung and
D.R. Nelson ({\em Phys. Rev A 38:1005 (1988)}), is interesting on its own and
allows to compute the energy for an {\em arbitrary} distribution of defects, 
on an {\em arbitrary geometry} with an arbitrary elastic {\em energy} with 
very minor additional computational effort. Some implications for recent 
experimental, computational and theoretical work are also discussed.

\end{abstract}

\vfill\newpage

\pagebreak

\section{Introduction}

Topological defects, mainly disclinations and dislocations, play a crucial role 
in the understanding of the physics of many two-dimensional (2D) systems \cite{NEL:01}. 
Although computational techniques such as analytical methods, 
Ewald summation techniques \cite{Ewald}, Monte Carlo simulations \cite{Strand:92}, etc..  
can be used in many different contexts, they become quite inefficient in other
situations, particularly with crystals having a boundary or lying 
on a curved background. In this paper, a new approach 
based on previous work by Seung and Nelson \cite{SN:88} will be discussed in detail. 
The computational methods developed allow to obtain the 
lowest energy configuration for an {\em arbitrary distribution of defects} in a crystal 
lying on an {\em arbitrary geometry} with an {\em arbitrary elastic energy} 
at zero temperature. The method is general enough to include Bravais lattices 
other than the triangular case. A discussion of the screening cloud of dislocations 
surrounding a disclination will be presented. This problem will be used as a testing 
ground where the results may be compared with
the analytical predictions from elasticity theory derived in this paper. 

\begin{figure}[ctb]
\centerline{\includegraphics[width=2.5 in]{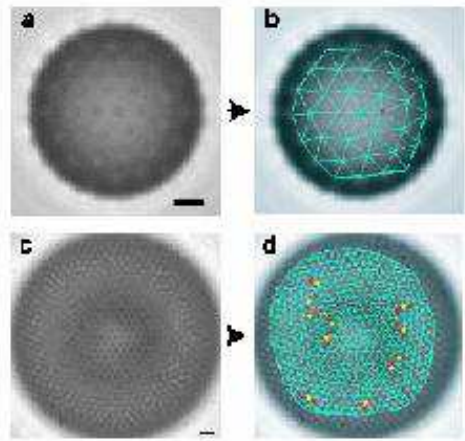}}
\caption{Ground state configurations for a large and small colloidosome. For a large 
colloidosome, finite length grain boundaries radiate out of the disclination 
(from \cite{Collo:03}).}
\label{fig__collo}
\end{figure}

An interesting experimental example where the dislocation cloud surrounding a disclination 
appear is given by colloidal particles crystallizing on the 
surface of a sphere (colloidosomes) \cite{Collo:03}. 
As a consequence of the Euler theorem \cite{Spi:79}, an sphere must 
have a total disclination charge of 12. If the total number of particles forming the
sphere is large enough, the ground state contains more defects than the 12 necessary
to satisfy the Euler theorem. Those additional defects are dislocations surrounding a
disclination, as illustrated in fig.~\ref{fig__collo}, arranging in the form of grain 
boundaries. Those grain boundaries have two very distinctive features 1)
terminate inside the medium and 2) have a total disclination charge of +1.
These features are intimately related to the geometry of the problem and should appear 
whenever the Gaussian curvature is large enough \cite{BNT:00,BowTra:01a,BCNT:02}.  
Similar structures have been also observed in simulations of the
Thomson problem \cite{PGDM:97,Alar,PGM:99}. Clouds of dislocations 
(with the minus disclinations playing the role of plus charges) should also appear in 
crystals on a negatively curved background \cite{GMS:82}. Those situations are relevant
as two dimensional analogs of the frustration 
associated with the three dimensional tetrahedral packing \cite{RuNe:83}
and other surface physics problems \cite{KlDo:85,SaCh:89}. 

The problem of grain boundaries radiating out of disclinations is also important
for understanding certain aspects of the problem of two-dimensional melting. 
The KTHNY \cite{KT,NH:79,Young} 
scenario predicts a two stage melting from a crystal into an isotropic phase via an
hexatic phase. This picture has been confirmed by a large number of examples
(see \cite{Strand:88,Strand:92} for reviews). Very precise 
numerical simulations \cite{SCKCM:97} have found very good agreement with the predictions
of the KTHNY scenario. Different inherent structures (IS) characterize each phase, and
disclinations appear as forming grain boundaries with very similar characteristics as the 
ones found in the Colloidosome problem. Similar structures have been found in recent simulations,
and the defects have been characterized by $1/f$ noise \cite{RR:03}. Recent experiments on plasmas \cite{QG:01} 
and colloids \cite{Tal:89,Mal:99}, have found also a similar situation, with correlation functions 
in good agreement with the
KTHNY scenario and grain boundaries with similar characteristics.
Alternative melting scenarios have exploited the short-range nature of the stresses produced
by grain boundaries of dislocations \cite{Chui:83}. 

The physics of Langmuir monolayers has received a lot of attention in recent years 
\cite{KD:92,KAG:99}, but many important questions have not yet been clarified. It seems very      clear that topological defects play a crucial role in problems such as melting or the collapse of
monolayers \cite{LKB:02}. More sophisitcated systems, such as sphingomyelin, where additional 
hydrogen bonding may be formed, show a much richer phase diagram \cite{VKO:02}. Biphasic
surfactant monolayers show additional defect structures (mesas) \cite{DWE:01}.

Radial grain boundaries radiating from a central disclination are also found in  
hexatic $I^{*}$ and crystal $J^{*}$ tilted liquid crystal phases \cite{DPM:86}. The tilt
is used to force a disclination and a pattern of radial grain boundaries (with 5 arms)
is observed.
 
The methods described in this paper are also relevant for a restricted type of quantum dots \cite{ZAPW:97}
in which the density of electrons is is small and the external disorder potential is weak enough
so that the electrons in the dot form a Wigner crystal, so called ``Wigner Crystal Islands'' \cite{KoShl:98}.
The methods presented in this paper are relevant to this case as well, as only a change of the boundary 
conditions for the problem is required.

Dislocation clouds also appear in many other problems 
such as, partially polymerized membranes \cite{RaNe:92,CN:93}, 
Wigner crystals \cite{FHM:79} and in carbon nanotubes, particularly
in the so called ``Onion'' rings, which are successive spherical layers of graphite
\cite{Slava}.

The organization of the paper is as follows. A review of continuum results is given in
Sect.~\ref{SECT__Cont}. The discretized approach is introduced in sect.~\ref{SECT__EFre} and
compared against the continuum results for isolated defects. 
The scaling relations satisfied by the energy are
derived in sect.~\ref{Scaling_Rel}. Numerical results are presented in
sect.~\ref{SECT__NUM} and their universality with respect to the Lame coefficients
is discussed in sect.~\ref{SECT__uni}. Sect~\ref{SECT__sph} is a brief overview of 
the effect of curvature. We wrap up with some conclusions in sect.~\ref{SECT__Con}. 
Several technicalities are relegated to the appendices.

\section{Continuum Results}\label{SECT__Cont}

The elastic energy of a continuum is given by \cite{Landau7}
\be\label{flat__spac}
F=\frac{1}{2}\int d^2 r (2\mu u^2_{\alpha \beta}+\lambda (u_{\alpha \alpha})^2)
\ee
where the strain tensor is defined as
\be\label{strain_flat__spac}
u_{\alpha \beta}=\frac{1}{2}(\frac{\parp{u_{\alpha}}}{\parp{x_{\beta}}}+
\frac{\parp{u_{\beta}}}{\parp{x_{\alpha}}}+
\frac{\parp{u_{\rho}}}{\parp{x_{\beta}}}\frac{\parp{u_{\rho}}}
{\parp{x_{\alpha}}}) \ .
\ee
The quadratic term in the strain tensor is usually dropped, as it usually amounts to
higher order negligible corrections to the total energy. Those terms cannot be entirely 
neglected if disclinations are involved \cite{SN:88} (For consistency, other 
energy terms quadratic in the strain tensor should also be included, but for the
sake of simplicity this point will be ignored). The Young modulus and Poisson ratio are
\bea\label{K_0sigma}
K_0&=&4\mu\frac{\mu+\lambda}{2\mu+\lambda} 
\nonumber\\ 
\nu&=&\frac{\lambda}{2\mu+\lambda} .
\eea

Using known analytical techniques, the strains for a dislocation and a 
disclination may be solved in linear order \cite{Landau7,SN:88}. For 
a dislocation with Burgers vector ${\bf b}=b {\bf e}_x$ the result is 
\cite{Landau7}
\bea\label{dlo_strain}
u_x&=&\frac{b}{2\pi}(\phi+\frac{K_0}{8\mu}\sin(2\phi))
\\\nonumber
u_y&=&-\frac{b}{2\pi}(\frac{\mu}{\lambda+2\mu}\log(r/a)+
\frac{K_0}{8\mu}\cos(2\phi)) .
\eea
The strains for disclinations of charge $s$ are \cite{SN:88}
\bea\label{dcl_strain}
u_r&=&-\frac{s}{2\pi}r(1-\frac{2\mu}{2\mu-\lambda}(A_s+1/2)-\frac{2\mu}{2\mu+\lambda}\log(r/R)) 
\nonumber\\
u_{\phi}&=&\frac{s}{2\pi}r\phi \ ,
\eea
There is an arbitrary constant $A_s$ which will be determined in the appendix sect.~\ref{APP_ZERO}.

Plugging Eq.~\ref{dlo_strain} and Eq.~\ref{dcl_strain} into Eq.~\ref{flat__spac}
The energies for a dislocation of Burgers vector $b$ is
\be\label{Sing_Disl}
F=\frac{|{\bf b}|^2}{8\pi}K_0(\ln(R/a)+\mbox{const}) .
\ee
As it is shown in the appendix sect.~\ref{APP_ZERO}, the constant $A_s$ is related
to the two-dimensional pressure $\Pi$. Following the same steps as for a dislocation, the
energy for an isolated disclination of charge $s=\frac{\pi}{3}q_i$ under pressure becomes
\be\label{Sing_Disclin}
F=\frac{s^2}{32 \pi}K_0 R^2+\frac{\pi \Pi^2}{8{\cal B}} R^2 \ .
\ee

\begin{figure}[ctb]
\centerline{\includegraphics[width=3in]{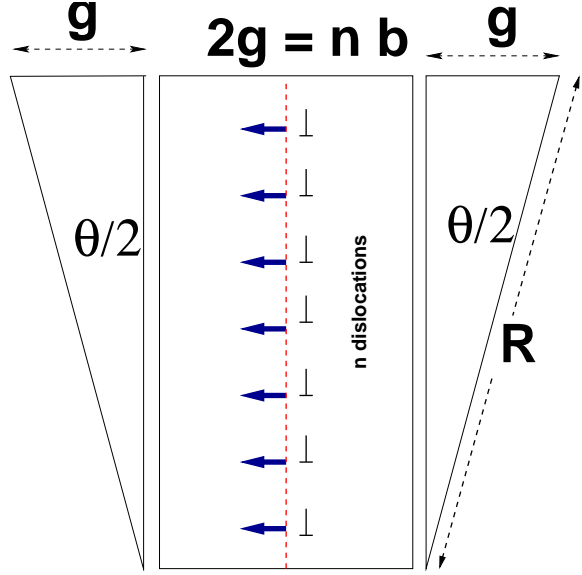}}
\caption{Relation between the aperture angle $\theta$ of a grain boundary of dislocations 
and its total Burgess vector $nb$ and radius of the crystal $R$.}
\label{fig__def}
\end{figure}

The energy dependence, growing as $R^2$, is energetically very costly.
Grain boundaries of dislocations may screen out the disclinations, thus reducing
the huge energy cost of an isolated disclination, if the angle 
of the grain boundary exactly compensates the missing or additional wedge 
(a multiple of $\frac{\pi}{3}$ for triangular
lattice) caused by the disclination. From the geometric argument in 
fig.~\ref{fig__def}, the angle of grain boundary $\theta$ is given by 
\be\label{spac__HH}
2R\sin(\theta/2)=nb ,
\ee
where $n$ is the total number of dislocations all having Burgers vector $b$. 
Further assuming a constant spacing of dislocations within the grain $D$, the
relation $B=Rb/D$ holds and one gets
\be\label{spac__apper}
2\sin(\theta/2)=b/D .
\ee
upon identifying $\theta$ with the missing or additional wedge removed to form the 
disclination \cite{BCNT:02}, Eq.~\ref{spac__apper} becomes
\be\label{spac__pred}
2\sin(\frac{\pi}{6m})=b/D \ ,
\ee
leading to an equation for the spacing $D$ as a function of the number of arms $m$.
A detailed analytical proof for this result using linear elasticity theory 
is provided in appendix~\ref{APP_ONE}, where the total energy of the system
of a disclination and an m-arm grain boundary (see fig.~\ref{GB_Characterist})
is given by Eq.~\ref{new_energy}. There is also a linear term in $R$ arising from 
two different contributions, the stresses of grain boundaries
of dislocations, discussed in appendix~\ref{APP__GB} and the core energies of the defects.
The total contribution to the energy is then,
\be\label{screen_energy}
F=\frac{(s-m \frac{b}{D})^2}{32 \pi}K_0 R^2 + \left(H(2\pi \frac{a}{D})+4\pi  c \right)  
m \frac{K_0}{4\pi} \frac{a^2}{D} R \ ,
\ee
where the function $H$ is defined in the appendix Eq.~\ref{H_d_gb}.
The pressure $\Pi$ has been set to zero. The total number of dislocations is
$n=m\frac{R}{D}$. The core energy has been parametrized as
$E_{core}=K_0 a^2 c$, where $c$ is a dimensionless coefficient (but not independent of
the elastic constants). 

If the spacing $D$ is given by Eq.~\ref{spac__pred}, the leading $R^2$ term in 
Eq.~\ref{screen_energy} is canceled and only the linear term in $R$ , which we denote as $f$, survives
\be\label{arg2}
f \equiv \left(H(2\pi \frac{a}{D})+4\pi c \right) m \frac{K_0}{4\pi} \frac{a^2}{D} R \ ,
\ee
For future reference, we quote the result for large $m$ 
\be\label{arg2_largem}
f=\left(\frac{1}{4\pi}(1-\log(\frac{2 \pi^2}{3m}))+ c\right)  m K_0 a^2\frac{R}{D} \ .
\ee
At perfect screening $m\frac{a}{D}=q_i \frac{\pi}{3}$ and
the energy grows logarithmically with $m$.
All previous results assume an infinite system (large $R$).
How large $R$ must be in order that the infinite radius result hold, will  
be estimated next.

The interaction energy of two grain
boundaries decays exponentially fast as a function of their mutual distance \cite{RS:50},
with a decay length 
\be\label{stres_out}
\lambda_{gb}=\frac{D}{2 \pi} ,
\ee
$D$ being the distance between dislocations within the grain (fig.~\ref{fig__def}).

dislocations that are a distance $L$ from
the disclination will be invisible to the dislocations in other arms if 
\be\label{arg3}
2  L \sin(\pi/m) >> 2 \lambda_{g} \ .
\ee
therefore from Eq.~\ref{stres_out} and Eq.~\ref{spac__pred} it follows 
\be\label{arg4}
L >> \frac{3}{\pi} m^2 b.
\ee
For $R>L$ the infinite result Eq.~\ref{arg2} will hold. The dependence of
Eq.~\ref{arg4} on $m^2$ points out that the behavior Eq.~\ref{arg2} for large number 
of arms will not be reached until the radius $R$ is very large.

\section{Discretization of the Elastic Free Energy}\label{SECT__EFre}

We consider a 2d triangular lattice of monomers with lattice constant $a$. 
The actual position of the $(n,m)$ monomer is described by ${\bf r}_{(n,m)}$
and it may be decomposed as
\be\label{dec_monomer}
{\bf r}_{(n,m)}=n {\bf e}_1+m{\bf e}_2+{\bf u}_{(n,m)} \ ,
\ee
where ${\bf u}_b$ is the strain at point $b$ and
${\bf e}_i, i=1,2$ define a basis of the Bravais lattice. For a 
triangular lattice it is 
\bea\label{tri_basis}
{\bf e}_1&=&{\bf e}_x
\\\nonumber
{\bf e}_2&=&\frac{1}{2}{\bf e}_x+\frac{\sqrt{3}}{2}{\bf e}_y \ .
\eea

The discrete elastic free energy that we will use is 
\be\label{Disc_Free}
F=\frac{\epsilon}{2}\sum_{<b c>}( |{\bf r}_{bc}|-1)^2 +
\sigma \sum_{b,c}(\frac{1}{2}-\frac{{\bf r}_{bc}\cdot{\bf r}_{bc+1}}
{|\bf{r}_{bc}||{\bf r}_{bc+1}|})^2 \ .
\ee
The summation in the first term runs over links defined by
vertices $a,b$, and in the second term the sum runs over all nearest 
neighbors of point $b$. Other discretizations are also possible but do
not modify the results as it will be discussed.

In order to proof the equivalence of Eq.~\ref{Disc_Free} with the continuum
result Eq.~\ref{flat__spac}, one must define the discrete derivatives first.
The derivatives of a general function $f({\bf r})$ are only defined along the
discrete direction of the triangular lattice, so a general derivative is 
precisely defined from
\bea\label{Disc_Deriv_x_y}
\parp_{x}f({\bf r}_b)&=&\frac{f({\bf r}_b+a{\bf e}_1)-f({\bf r}_b)}{a}
\\\nonumber
\parp_{y}f({\bf r}_b)&=&\frac{f({\bf r}_b+a{\bf e}_2)-f({\bf r}_b)
-\frac{1}{2}(f({\bf r}_b+a{\bf e}_1)-f({\bf r}_b))}{a\sqrt{3}/2} \ .
\eea
Using the previous definition, the discrete metric becomes
\be\label{Disc_metric}
g_{i j}(b)=\parp_{i} {\bf r}_b \parp_{j}{\bf r}_b \ .
\ee
The distance between two nearest neighbor points is
\be\label{disc_distance}
|{\bf r}_a-{\bf r_b}|=(1+2 u_{i j}e^i_{ab}e^j_{ab})^{1/2} \ ,
\ee
where the discrete strain tensor is defined from the derivatives 
Eq.~\ref{Disc_Deriv_x_y} by
\be\label{strain_flat__spac_disc}
u_{\alpha \beta}=\frac{1}{2}(\frac{\parp{u_{\alpha}}}{\parp{x_{\beta}}}+
\frac{\parp{u_{\beta}}}{\parp{x_{\alpha}}}+
\frac{\parp{u_{\rho}}}{\parp{x_{\beta}}}\frac{\parp{u_{\rho}}}
{\parp{x_{\alpha}}}) \ .
\ee
Expanding Eq.~\ref{disc_distance} using Eq.~\ref{app_four_a} derived in the appendix, the first
term of the discrete energy is identical to Eq.~\ref{flat__spac} with
\bea\label{First_Term}
\lambda&=&\frac{\sqrt{3}}{4} \epsilon  
\nonumber\\
\mu&=&\frac{\sqrt{3}}{4} \epsilon \ .
\eea
The same expansion to the second term using the relations
Eq.~\ref{app_four_b}, \ref{app_four_c}, \ref{app_four_d} from the appendix lead to
\bea\label{Second_Term}
\lambda&=&9\frac{\sqrt{3}}{4} \sigma
\nonumber\\
\mu&=&-9\frac{\sqrt{3}}{4} \epsilon \ .
\eea
The elastic constants of the two terms combined are then
\bea\label{Two_Terms}
\lambda&=&\frac{\sqrt{3}}{4}(\epsilon+9\sigma)  
\nonumber\\
\mu&=&\frac{\sqrt{3}}{4}(\epsilon-9\sigma)  \ ,
\eea
and therefore, Eq.~\ref{Disc_Free} provides a suitable discretization of
Eq.~\ref{flat__spac} with arbitrary elastic constants. It should be recalled that
higher order terms in the displacement are dropped to lead to Eq.~\ref{Two_Terms}.  

In this paper, the particular geometry that will be used consists of
a plus or minus disclination (a pentagon or an heptagon respectively) at the center of
the crystal. The total number of monomers forming the crystal is a
function of the linear size $R$ and depends on the total number of dislocations.
If only a center defect of charge $q_i$ is present, the total number of monomers is
\be\label{tot_number}
M=(6-q_i)\frac{R^2+R}{2}+1 \ ,
\ee
Even with additional dislocations, 
the total number of points will still grow like $(6-q_i)R^2$.

\subsection{Energies of single defects}

\begin{figure}[ctb]
\centerline{\includegraphics[width=0.9\hsize]{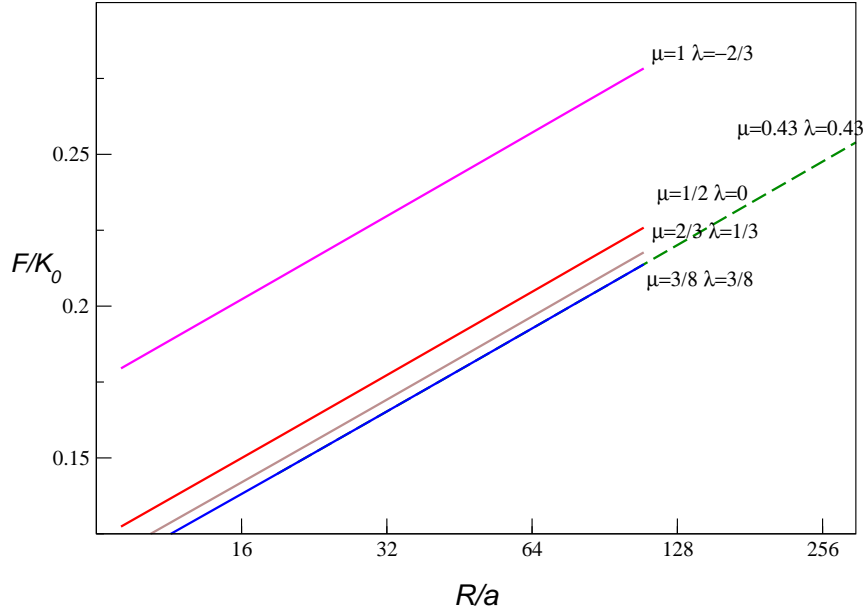}}
\caption{Results for the energy of an isolated dislocation as function of R
for different values of the elastic constants.}
\label{fig__Sing__Disl}
\end{figure}

The energies of isolated dislocations and disclinations will be 
compared against the analytical predictions derived previously. This will provide a
benchmark to the present approach.

In fig.~\ref{fig__Sing__Disl} the minimum energy results
of a configuration containing a dislocation for different values of the
Lame coefficients. A fit to the form of Eq.~\ref{Sing_Disl} yields
\be\label{sing_disl}
\frac{F}{K_0 |{\bf b}|^2}=0.0395(2)\log(r)
\ee
in agreement within of $1\%$ with the analytical result 
$\frac{1}{8\pi}=0.03979$. The small deviation may be attributed to the neglect
of higher order terms in the continuum calculation. Core energies may be computed from
the intercept in fig~\ref{fig__Sing__Disl}. The results are summarized in 
table~\ref{Tab__core}. It should be noted that core energies for crystals 
with the same Young modulus are different.  

\begin{table}[ctb]
\centerline{
\begin{tabular}{|c||c|c|c|c|c|}\hline
 $(\mu,\lambda)$   & $(1,-\frac{2}{3})$ & $(\frac{1}{2},0)$ & 
 $(\frac{2}{3},\frac{1}{3})$ & $(\frac{3}{8},\frac{3}{8})$ & $(\frac{\sqrt{3}}{4},
  \frac{\sqrt{3}}{4})$ \\\hline 
 $\frac{E_{core}}{K_0 a^2}$ & $0.092(1)$ &$0.042(1)$ &$0.039(1)$ & $0.029(1)$ &$0.0285(1)$ 
\\\hline
\end{tabular}}
\caption{Core energies of dislocations as a function of Lame coefficients $(\mu,\lambda)$}
\label{Tab__core}
\end{table}

The results of the same analysis for single disclinations are shown in table~\ref{Tab__coeff}.
The most accurate determination yields
\be\label{ee}
\frac{F}{R^2 K_0 s^2}=0.00785(1) \ ,
\ee
both for plus and minus disclinations (Results for $\lambda=\mu$ were first obtained 
with less accuracy by \cite{SN:88}). Although the coefficient is off
from the analytical result Eq.~\ref{Sing_Disclin} by a significant amount,
it is remarkably universal as a function of the elastic constants.
The energies are also the same for both plus and minus disclinations 
within a $0.1\%$ accuracy (more accurate results do show that sevenfold defects 
have a marginally lower energy). There is, however, a term that grows linearly 
with $R$ in the discretized energy. This term is quoted also in table~\ref{Tab__coeff}.

\begin{figure}[ctb]
\centerline{\includegraphics[height=0.35\hsize]{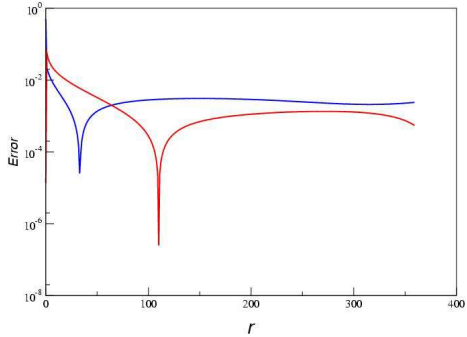}}
\caption{Plot of function $E(r_b)$ describing the relative difference of the 
analytical solution and the numerical result.}
\label{Discl_error}
\end{figure}

\begin{table}[ctb]
\centerline{
\begin{tabular}{|c||c|c|c|c|}\hline
 $s$   & $(1,-\frac{2}{3})$ & $(\frac{1}{2},0)$ & $(\frac{2}{3},\frac{1}{3})$ & 
 $(\frac{3}{8},\frac{3}{8})$ \\\hline 
 $R^2 (+)$ & $0.80(1)$  &$0.79(1)$ &$0.078(1)$   & $0.0785(1)$ \\\hline
 $R^2 (-)$ & $0.79(1)$  &$0.78(1)$ &$0.079(1)$   & $0.0785(1)$ \\\hline
 $R   (+)$ & $-0.003(1)$  &$0.006(1)$&$0.007(1)$ & $0.08(1)$ \\\hline
 $R   (-)$ & $-$  &$0.007(1)$ &$0.008(1)$   & $0.009(1)$ \\\hline
\end{tabular}}
\caption{Coefficient (Eq.~\ref{ee}) as a function of the Lame coefficients $(\mu,\lambda)$}
\label{Tab__coeff}
\end{table}

It is also instructive to compare the continuum strain solution ${\bf r}_A$
Eq.~\ref{dcl_strain} with the configuration from the discrete calculation.
The difference will be quantified from the function
\be\label{err_rat}
E(r_b)=\frac{|{\bf r}_b-{\bf r}_A|}{|{\bf r}_b|+|{\bf r}_A|}
\ee
and it is plotted in fig.~\ref{Discl_error}. The relative error is always below
$0.01\%$, becoming as low as $10^{-6}\%$. 

\section{Scaling Relations for Grain Boundaries}\label{Scaling_Rel}

\begin{figure}[ctb]
\centerline{\includegraphics[height=0.55\hsize]{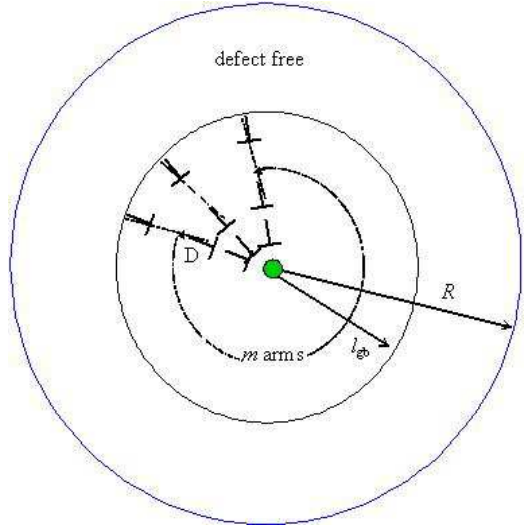}}
\caption{Degrees of freedom of a grain boundary.}
\label{GB_Characterist}
\end{figure}

The degrees of freedom defining a grain boundary of dislocations are
(see fig.~\ref{GB_Characterist}):
\begin{itemize}
\item{$m$: number of arms.}
\item{$l_{gb}$: Length of the grain boundary from the central disclination.}
\item{$\Psi$: angle of the grain with some specified crystallographic axis.}
\item{$D$: spacing of dislocations within each grain.}
\end{itemize}
The spacing $D$ does not generally need to be restricted to be constant. The
linear size of the system is $R$.

It is convenient to define the dimensionless variable
\be\label{scal_vari}
x=l_{gb}/R \ ,
\ee
which by definition satisfies $0\le x \le 1$. $x=1$ implies that grains
reach the boundary while $x=0$ implies no dislocations. 
Eq.~\ref{screen_energy} will be generalized (for zero temperature) by assuming that 
for intermediate values of $x$ the following scaling behavior holds
\be\label{energy_x}
E=K_0 R^2 {\cal Q}(x,m,D,\Psi)  
\ee
where as $x \rightarrow 0$ ${\cal Q}$ approaches Eq.~\ref{Sing_Disclin}. This scaling
law however, may break down if the sub-leading linear terms in $R$ become comparable. 
The condition expressing this situation is given by (see Eq.~\ref{arg2}).
\be\label{bal_energy}
{\cal Q}(x_c) \sim \frac{\pi x_c}{3 R} f \ ,
\ee
If the term square in $R$ vanishes,
this equation must hold for some $x_c$ and scaling will break down for all $x>x_c$. 

If the square term in $R$ does not cancel,
then the scaling relation must hold for large $R$ as well. 
That will be the case for  spacings $D$ not satisfying 
Eq.~\ref{spac__pred}. The ${\cal Q}$ function
should exhibit a minima roughly at the critical $x_m$ where the additional 
angle added by the dislocations compensates the missing or additional angle
by the disclinations. The critical $x_m$ is given by 
\be\label{crit}
x_m=2 D \sin(\pi/(6m)) \ ,
\ee 
with the additional constraint $x_m<1$. Therefore the ${\cal Q}$ should exhibit a 
minima for $x_m$.

The angle $\Psi$, defined as the angle of the grain with respect
a crystallographic axis, for an m-grain boundary commensurate with the 
p-fold symmetry of the central disclination will be constrained to
\be\label{angle_all}
0 \le \Psi < \frac{\pi}{m} \ .
\ee
A straight-forward analytical calculation shows that ${\cal Q}$ should be
independent of $\Psi$. It will be shown that this result ignores the constraints imposed
on the Burgers vector by the underlying lattice.

\section{Numerical study of grain Boundaries}\label{SECT__NUM}

\subsection{Some computational details}

The calculations have been done by relaxing an initial configuration
consistent with the given distribution of defects using the conjugate
gradient method. The numerical accuracy was tested by checking the convergence of
the final results as a function of the tolerance error in the algorithm.
Whenever different initial configurations consistent with the given distributions of defects 
were tried, the final result was found 
to be identical. The energies were computed to a seven digit precision or more.

For each value of the parameters, results were obtained for linear sizes ranging from
$R=10$ to $R=200$ corresponding to typical volume sizes from $247$ to $140000$ monomers.
Those lattices are, in many cases, larger than some experimental systems available. Free boundary
conditions were incorporated by allowing the system to reach its natural extend without any
external constraint.

The code has been written in C++ using objected oriented design. This provides the flexibility of
incorporating any potentially new geometry, distribution of defects or discretization energy
at any future time with a very minute effort. The code has some limitations too, one of them
being that non-integer spacings (cases were 
dislocations within the grain should be slightly non-constant to mimic 
fractions of the lattice spacing) have not
been implemented effectively. Typical minimizations of a system with 
$R=100$ (total number of monomers $M=25000$) take 9 minutes at a $10^{-10}$ precision
and 5 minutes at $10^{-5}$ precision in a Dell 1.80 GHz dual 
Xeon procesor running Linux Red Hat. Further computational details will be 
presented elsewhere.

\subsection{Scaling as a function of $l_{gb}$ and $D$}

A plus disclination will be placed at the center of the crystal and grain boundaries
will be fixed to have angle $\Psi=\pi/5$ and number of arms 
$m=5$. The energy will be investigated as a function of the parameters $l_{gb}$ and $D$. 
The elastic constants will be chosen as $\lambda=\mu$.

\begin{figure}[hp]
\centerline{\includegraphics[height=.6\hsize]{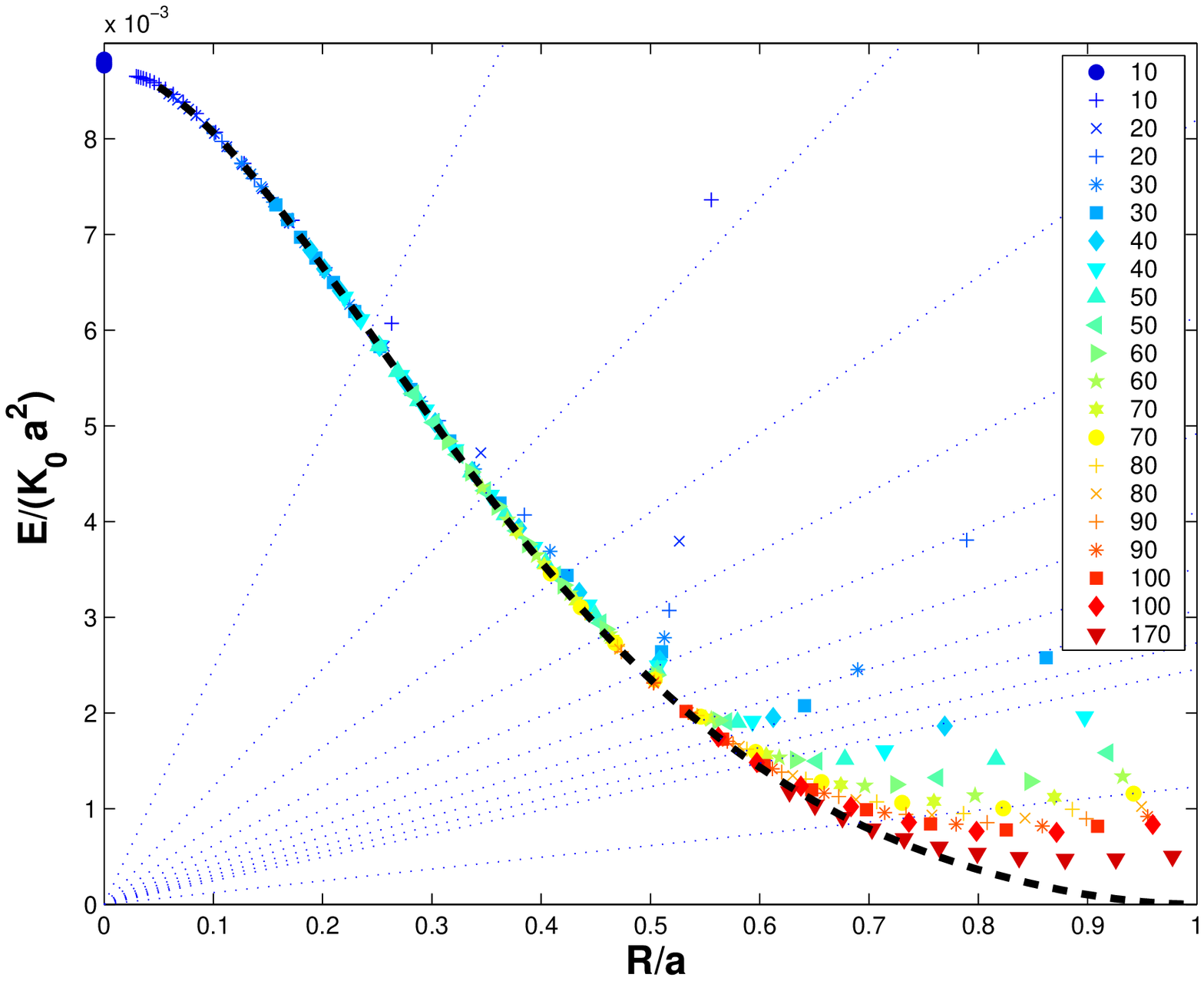}}
\caption{Plot of the scaling function Eq.~\ref{energy_x} for $D=5$, $m=5$, $\Psi=\pi/5$ and
sizes $R=10-170$. The dashed line is a fit with an extrapolation to $x->1$.
The straight lines provide a visual solution to Eq.~\ref{bal_energy} and an estimate 
of the critical $x_c$ as a function of $R$.}
\label{fig__scaling__spac5__m5}
\centerline{\includegraphics[height=.6\hsize]{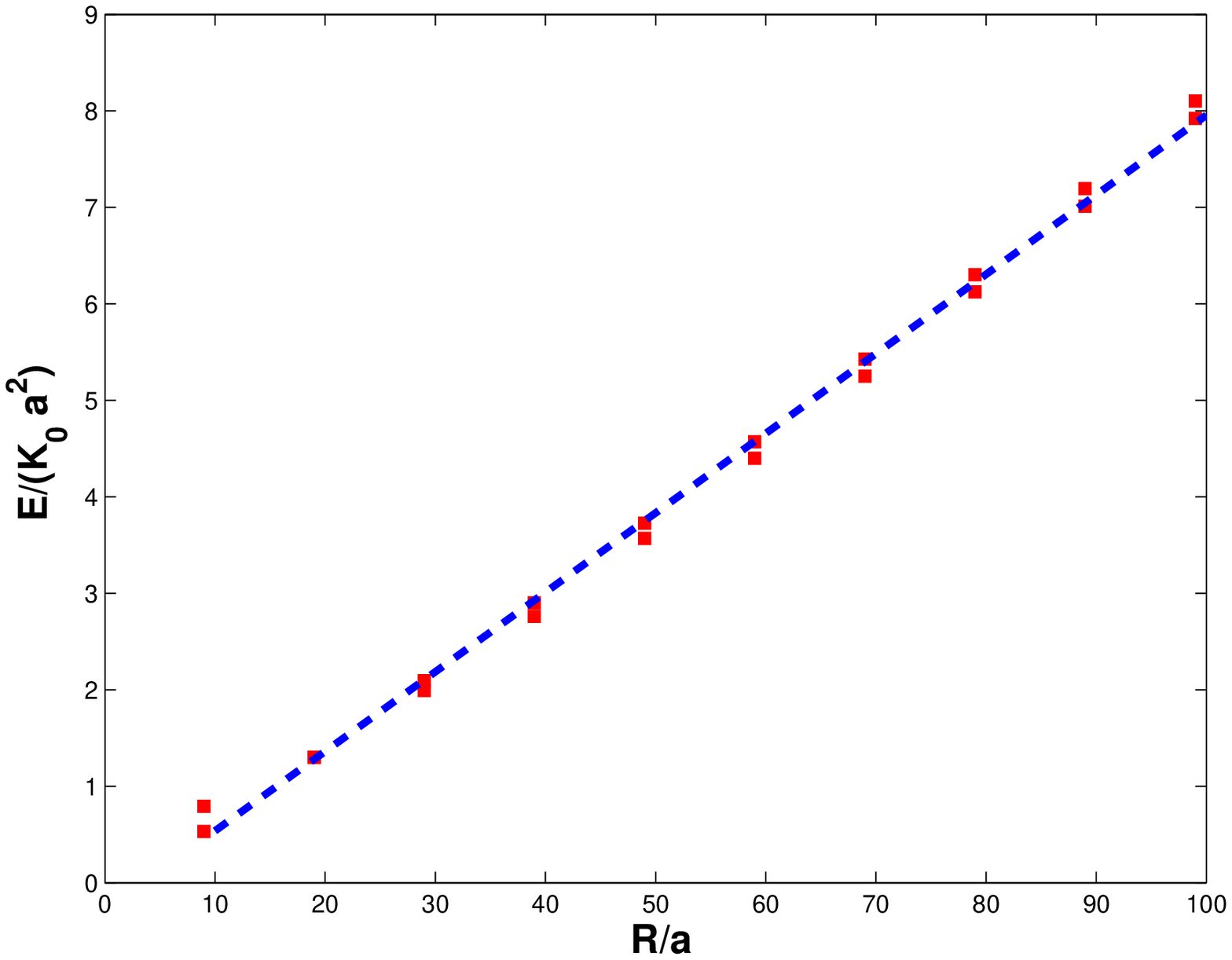}}
\caption{Plot of the fit Eq.~\ref{arg2} to the results for $R\sim l_{gb}$.}
\label{fig__noscaling__fit}
\end{figure}

The energy of relaxed configurations is plotted in fig.~\ref{fig__scaling__spac5__m5}
as a function of the scaling variables. the ${\cal Q}$ function is plotted
for a plus-disclination and a $m=5$ grain boundary with spacing $D=5$. The plots
follow very well the ansatz in Eq.~\ref{energy_x}, with obvious deviations for
values of $x$ closer to 1.

In order to determine the coefficient of the linear term in R, defined as $f$ in 
Eq.~\ref{arg2}, it will be assumed that the points on fig.~\ref{fig__scaling__spac5__m5} 
for $x \sim1$, which clearly do not scale, are described by Eq.~\ref{arg2}. The results 
are plotted in fig.~\ref{fig__noscaling__fit} and one obtains for the $f$-coefficient
\be\label{fit_depen}
f=0.081(5) \ .
\ee
Since we are not looking for a high precision value for the $f$-coefficient, two values for
$l_{gb}$ were included for any given $R$. This provides strong evidence of the robustness
of the fit. The theoretical value of the $f$ coefficient arises from a
grain boundary contribution Eq.~\ref{fin_en_d_gb} and a core energy which may be estimated
from a single dislocation (given in table~\ref{Tab__core}). The two contributions give
\be\label{fit_depen_theo}
f= 0.063 \mbox{(Grain Boundary)}+0.029\mbox{(Core Energy)}=0.092 \ ,
\ee
considering the approximations involved (basically linear elasticity theory and
non-interacting grains), the agreement with Eq.~\ref{fit_depen} is acceptable.

The value obtained for $f$ may be cross-checked by assuming that the scaling behavior
will break down when the value for the linear term is, say one third, of the square term.
The intersection of the straight lines in fig.~\ref{fig__scaling__spac5__m5} provide
a visual solution to Eq.~\ref{bal_energy} and an estimate of the critical $x_c$ as a function
of $R$ in reasonable self-consistent agreement, particularly for large values of $R$.

Three regions may be clearly identified from fig.~\ref{fig__scaling__spac5__m5}:
\begin{itemize}
\item{($x \sim 0$): The energy is dominated by the strains of the central disclination and 
the effect of adding additional defects is negligible.}
\item{($0<< x < x_c$): The inclusion of more defects dramatically lowers the energy.}
\item{($x > x_c$): The core energy contribution sets in and scaling breaks down, the energy 
grows linearly with $R$, as apparent from fig.~\ref{fig__noscaling__fit}.}
\end{itemize}
It should be noted from fig.~\ref{fig__scaling__spac5__m5} that if the intermediate region 
could be extrapolated to to $x \sim 1 $ before the core energy terms would become 
noticeable, the energy would go to zero as $(x-1)^2$, that is, independent of the linear
size $R$.

A typical relaxed final configuration is shown in fig.~\ref{fig__5_5}. The plot is only for
a small region around the central disclination, but it clearly shows that for regions 
away of the defects the triangles forming the crystal are almost equilateral (unstrained).

\subsection{Scaling function for non-optimal $D$ values at fixed $m$ and $\Psi$}

\begin{figure}[hp]
\centerline{\includegraphics[height=.6\hsize]{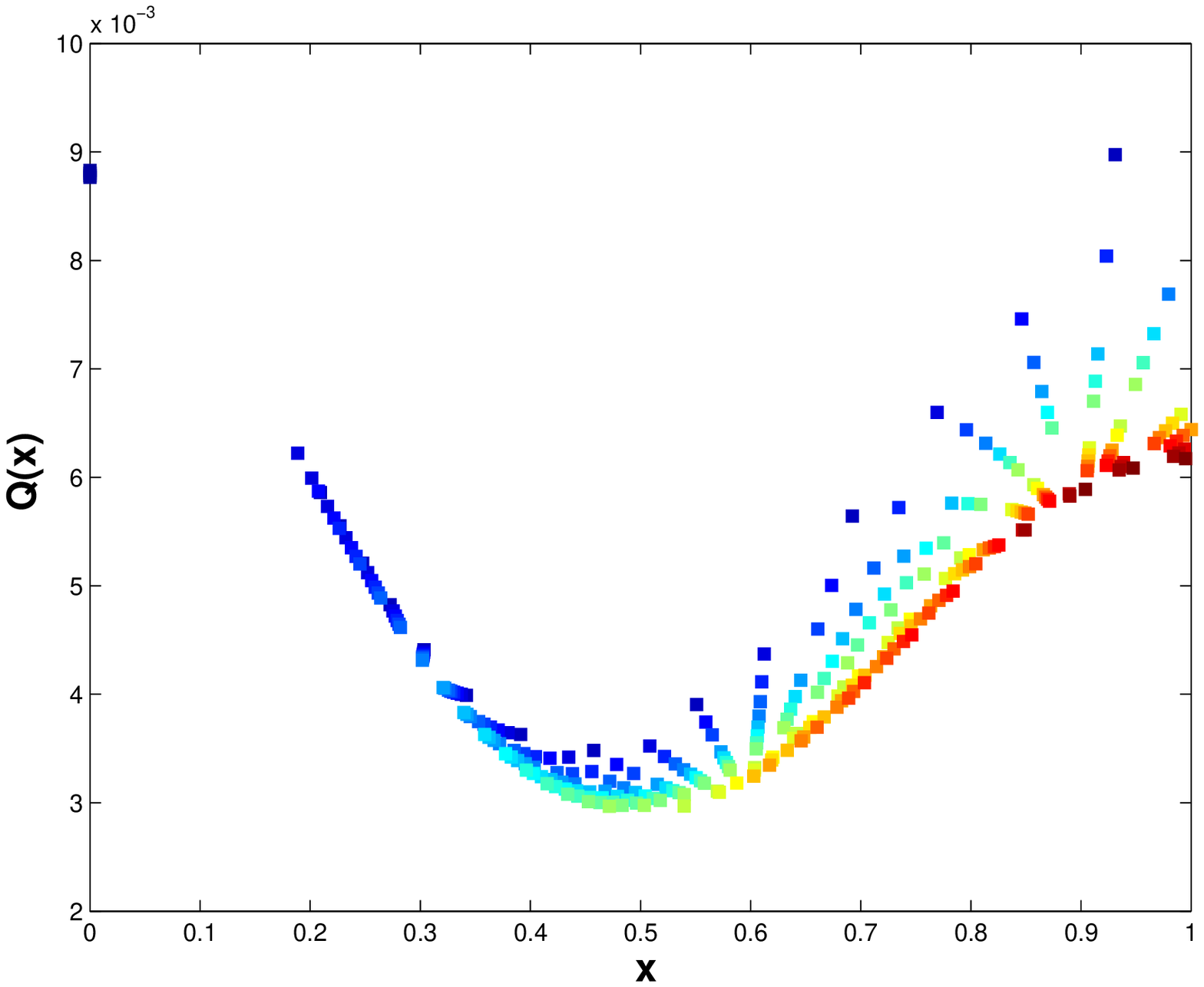}}
\caption{Plot of the scaling function Eq.~\ref{energy_x} for $D=3$, $\lambda=\mu$, 
$m=5$ and $\Psi=\frac{\pi}{5}$ for sizes $R=10-200$ ($\lambda=\mu$).}  
\label{fig__scaling__spac3__m5}
\centerline{\includegraphics[height=.6\hsize]{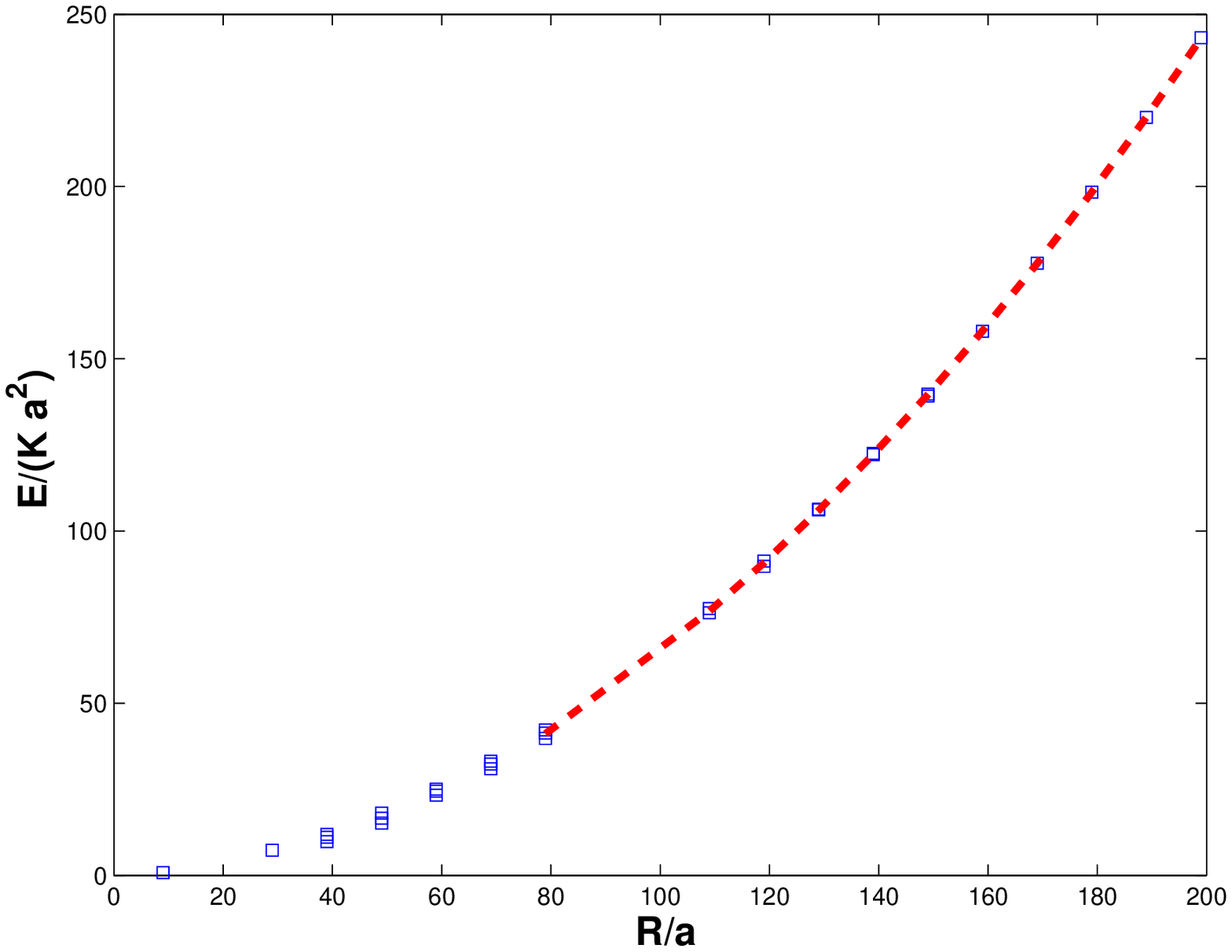}}
\caption{fit Eq.~\ref{screen_energy} to the values $ R \sim {l_{gb}}$.}
\label{fig__scaling__spac7__m5_fit}
\end{figure}

\begin{figure}[hp]
\centerline{\includegraphics[height=.5\hsize]{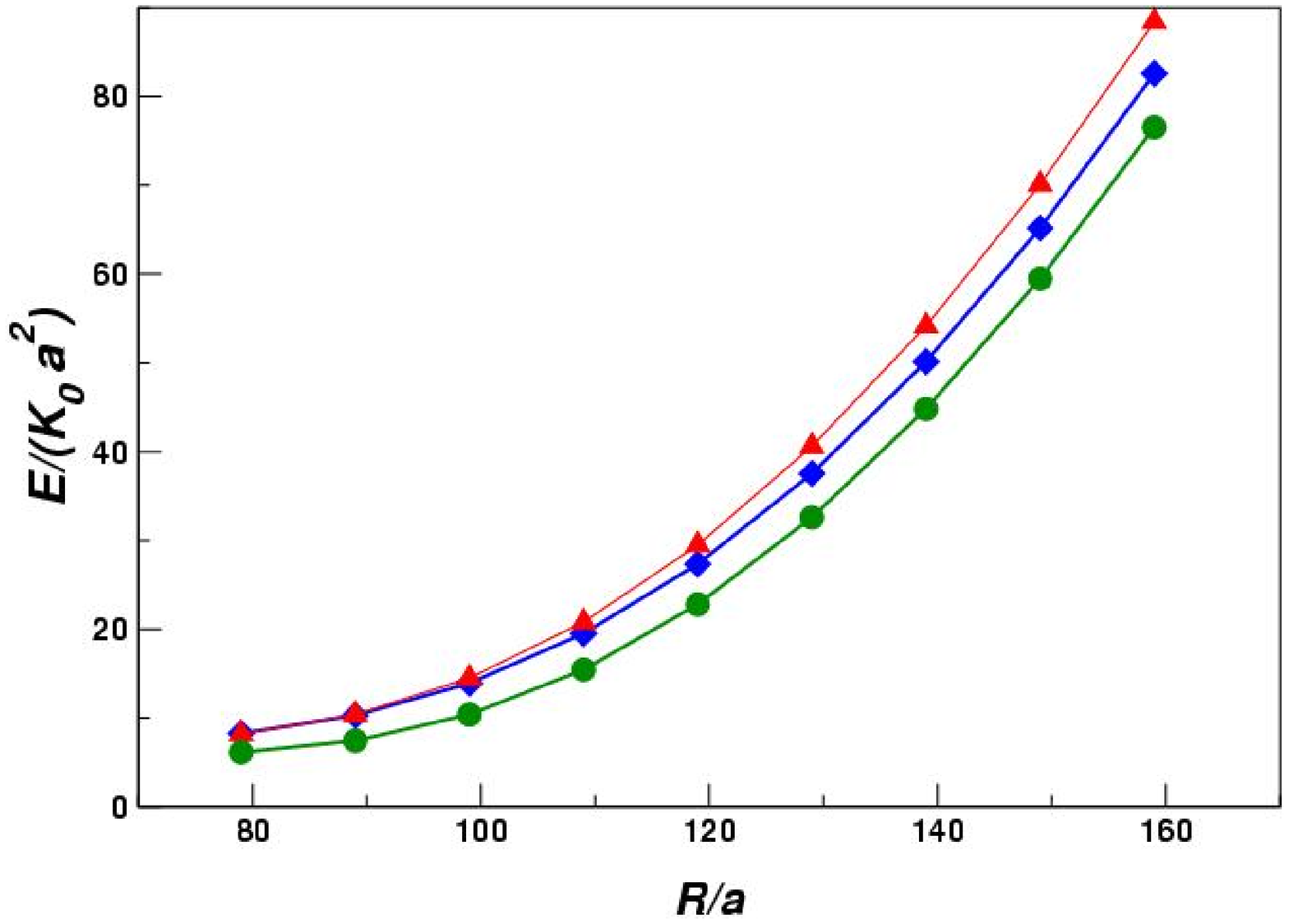}}
\caption{Comparison of different energies for $m=5$ and $D=5$ $\lambda=\mu$ for different
values of $\Psi$.}
\label{fig__m5_diff_angl}
\centerline{\includegraphics[height=.6\hsize]{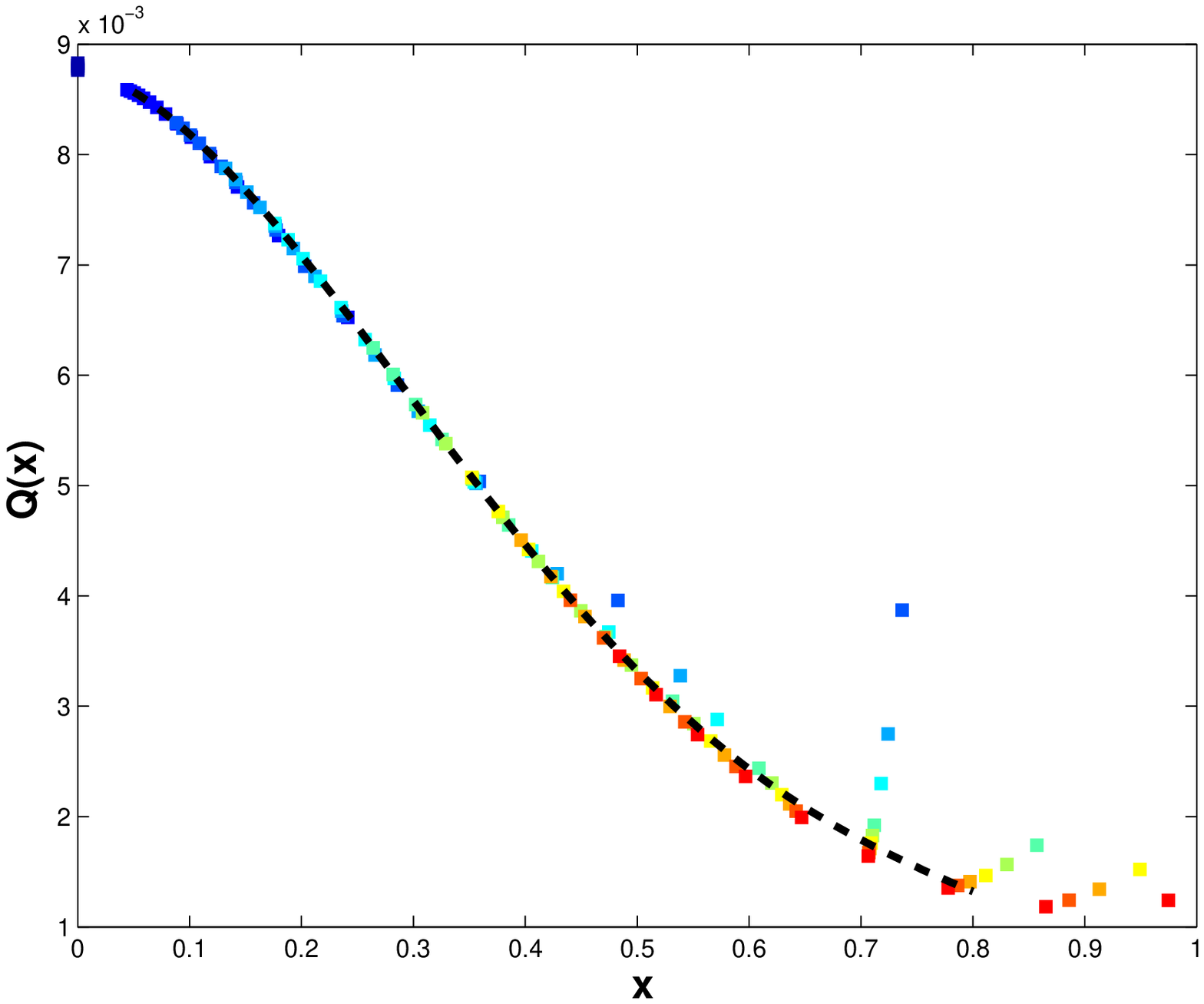}}
\caption{Plot of the scaling function Eq.~\ref{energy_x} for $D=7$, $\lambda=\mu$, 
$m=5$ and $\Psi=\frac{\pi}{5}$ for sizes $R=10-100$ ($\lambda=\mu$).}  
\label{fig__scaling__spac7__m5}
\end{figure}

The ${\cal Q}$ function for $D$ values significantly different than 
$D_{crit}=\frac{5}{3}\pi$ are plotted in fig.~\ref{fig__scaling__spac3__m5} ($D=3$) and
fig.~\ref{fig__scaling__spac3__m5} ($D=7$). 

For $D<D_{crit}$ the scaling function has a minima as a function of $x$ in 
reasonable agreement with the value predicted by Eq.~\ref{crit}. As
it is very apparent from fig.~\ref{fig__scaling__spac7__m5_fit}
the energy grows quadratically (compare with fig.~\ref{fig__scaling__spac7__m5_fit})
with the system size, even in the presence of the grain boundaries. The fit gives
\be\label{d3_fit}
\frac{E}{K_0 R^2}=0.0055(4) \ .
\ee
This result should be compared with the theoretical estimate
Eq.~\ref{new_energy} for $\Pi=0$, $D=3$, $m=5$
\be\label{d3_fit_theo}
\frac{E}{K_0 R^2}=0.0039 \ .
\ee
Although the result is not too different, the quantitative agreement 
is not very good. This may be attributed to assuming linear elasticity, which as
already seen is not very accurate if disclinations are involved.

The energy for $D=7$ does not exhibit a minima because that should appear
for values of $x>1$. The total energy for $D=7$ is significantly larger than 
for $D=5$, as it becomes apparent from the relative scale of the y-axis of the plot. 
The theoretical estimate Eq.~\ref{new_energy} for $\Pi=0$, $D=7$, $m=5$ is
\be\label{d7_fit_theo}
\frac{E}{K_0 R^2}=0.0011 \ .
\ee
The result in fig.~\ref{fig__scaling__spac7__m5} approaches this limit, although 
the statistics are not as good as in for $D=3$.

\subsection{Dependence on the orientation of the grain $\Psi$}\label{subsect__or}

\begin{figure}[hp]
\centerline{\includegraphics[width=4in]{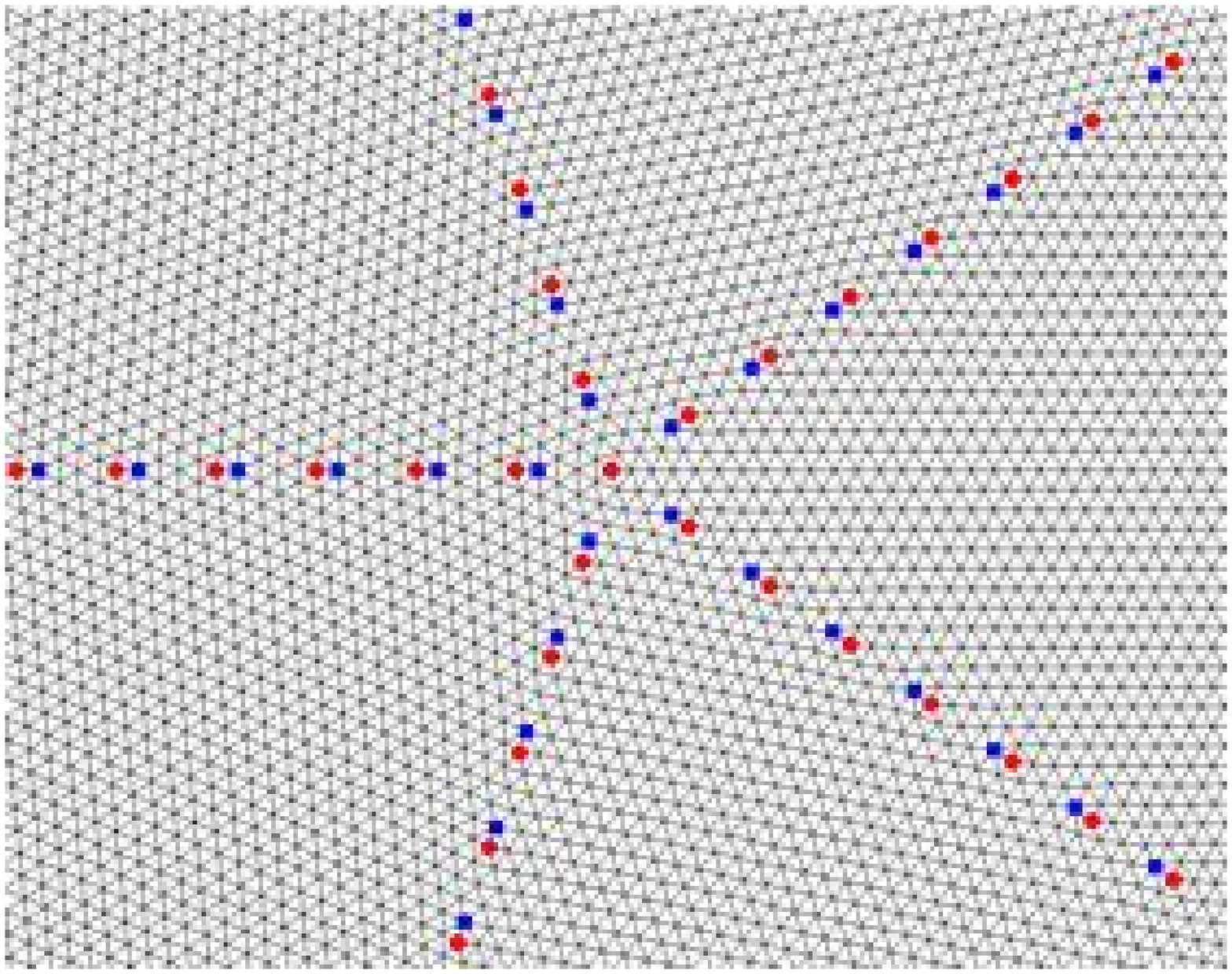}}
\caption{Final configurations for $m=5$, $D=5$ and $\Psi=\frac{\pi}{5}$ and $\lambda=\mu$.
Results correspond to $R=90$.}
\label{fig__5_5}
\centerline{\includegraphics[width=4in]{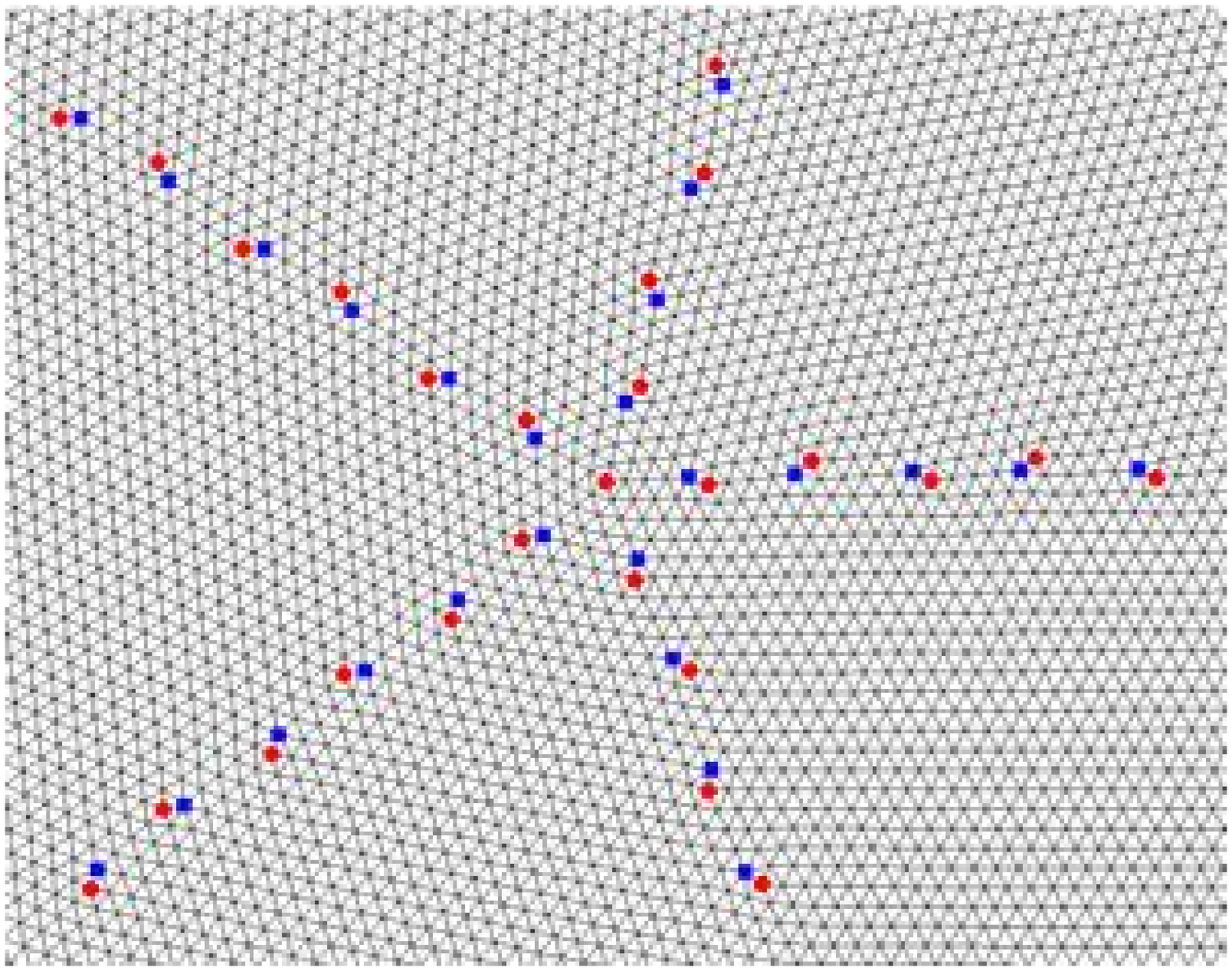}}
\caption{Final configurations for $m=5$, $D=5$ and $\Psi=0$ and $\lambda=\mu$.
Results correspond to $R=90$.}
\label{fig__5_5_ang0}
\end{figure}

The dependence on the angle $\Psi$ is plotted in fig.~\ref{fig__m5_diff_angl} for angles
$\Psi=0,\frac{\pi}{10},\frac{\pi}{5}$. Typical final relaxed configurations are shown in
fig.~\ref{fig__5_5} ($\Psi=\pi/5$) and fig.~\ref{fig__5_5_ang0} ($\Psi=0$). The 
final relaxed configurations for $\Psi=\frac{\pi}{5}$ are very regular while the dislocations
for the angle $\Phi=0$ display a rather jagged pattern.  
Provided that the total Burgers vector is zero, the only difference in the energies
arises from the grain boundary terms. Therefore, the energy
difference in the results fig.~\ref{fig__m5_diff_angl} should be attributed to the constraints
induced by the lattice to the Burgers vector.This point might be substantiated 
numerically with a more comprehensive calculation, but this has not been done, partly
because the angle $\Psi$ is not well defined for values of $m$ different than 5.
The previous considerations reflect that the optimal angle for $m=5$ grain boundaries
is given by $\Psi=\pi/5$.

\subsection{Dependence on the number of arms $m$}

\begin{figure}[hp]
\centerline{\includegraphics[width=4in]{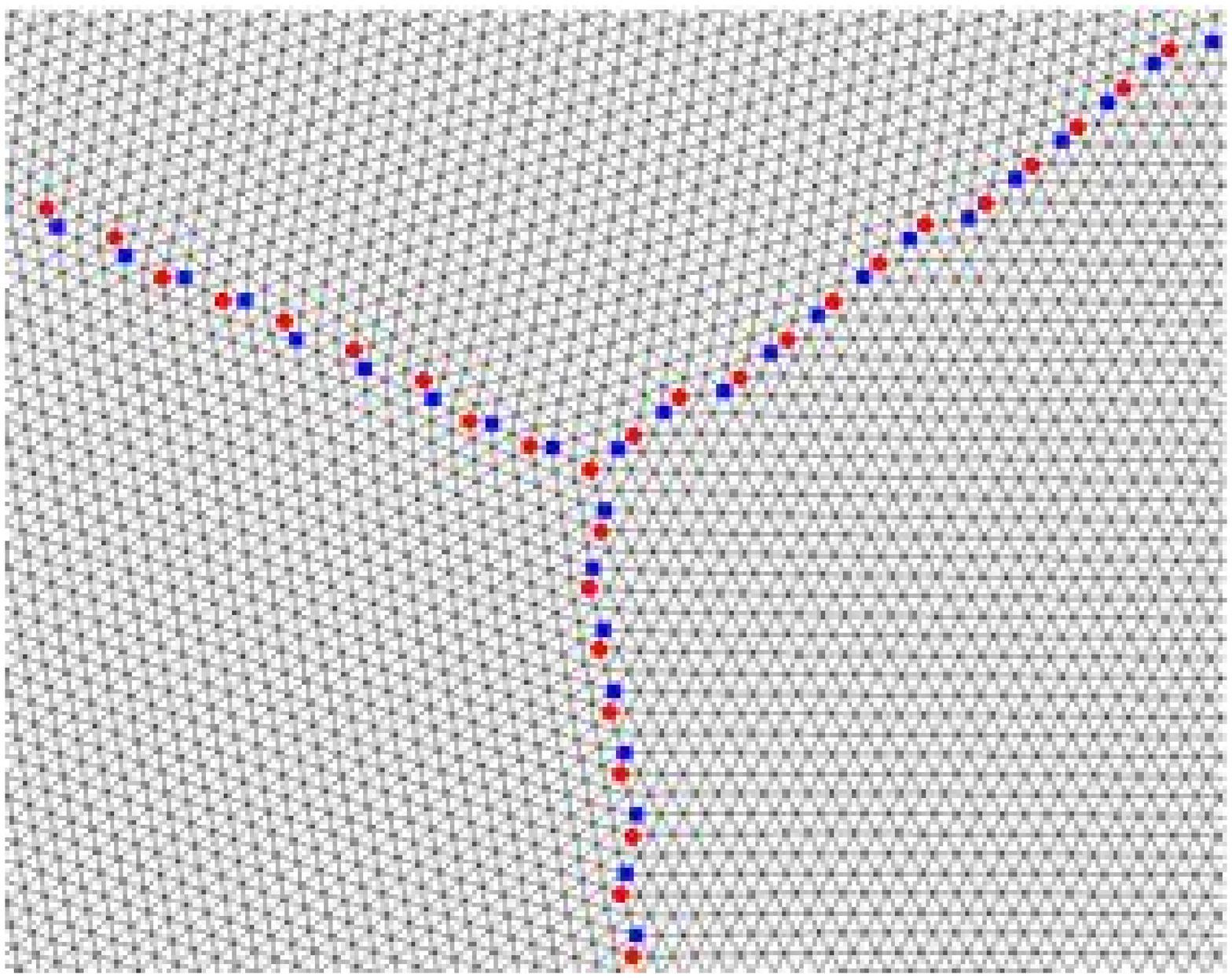}}
\caption{Relaxed configurations for $m=3$, $D=3$ and $\lambda=\mu$.
Results correspond to $R=90$.}
\label{fig__3_3}
\centerline{\includegraphics[width=4in]{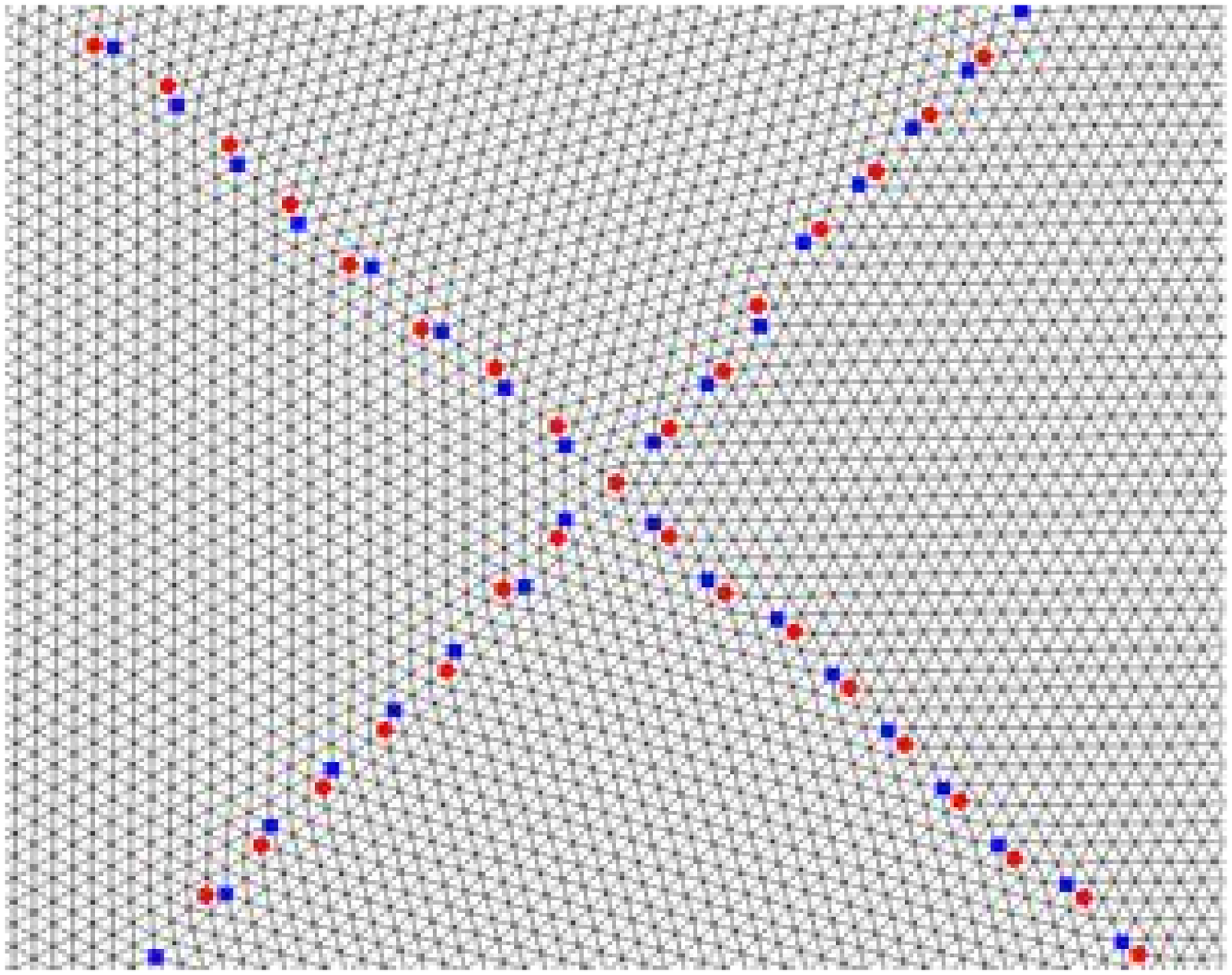}}
\caption{Relaxed configurations for $m=4$, $D=4$ and $\lambda=\mu$.
Results correspond to $R=90$.}
\label{fig__4_4}
\end{figure}

\begin{figure}[hp]
\centerline{\includegraphics[width=4in]{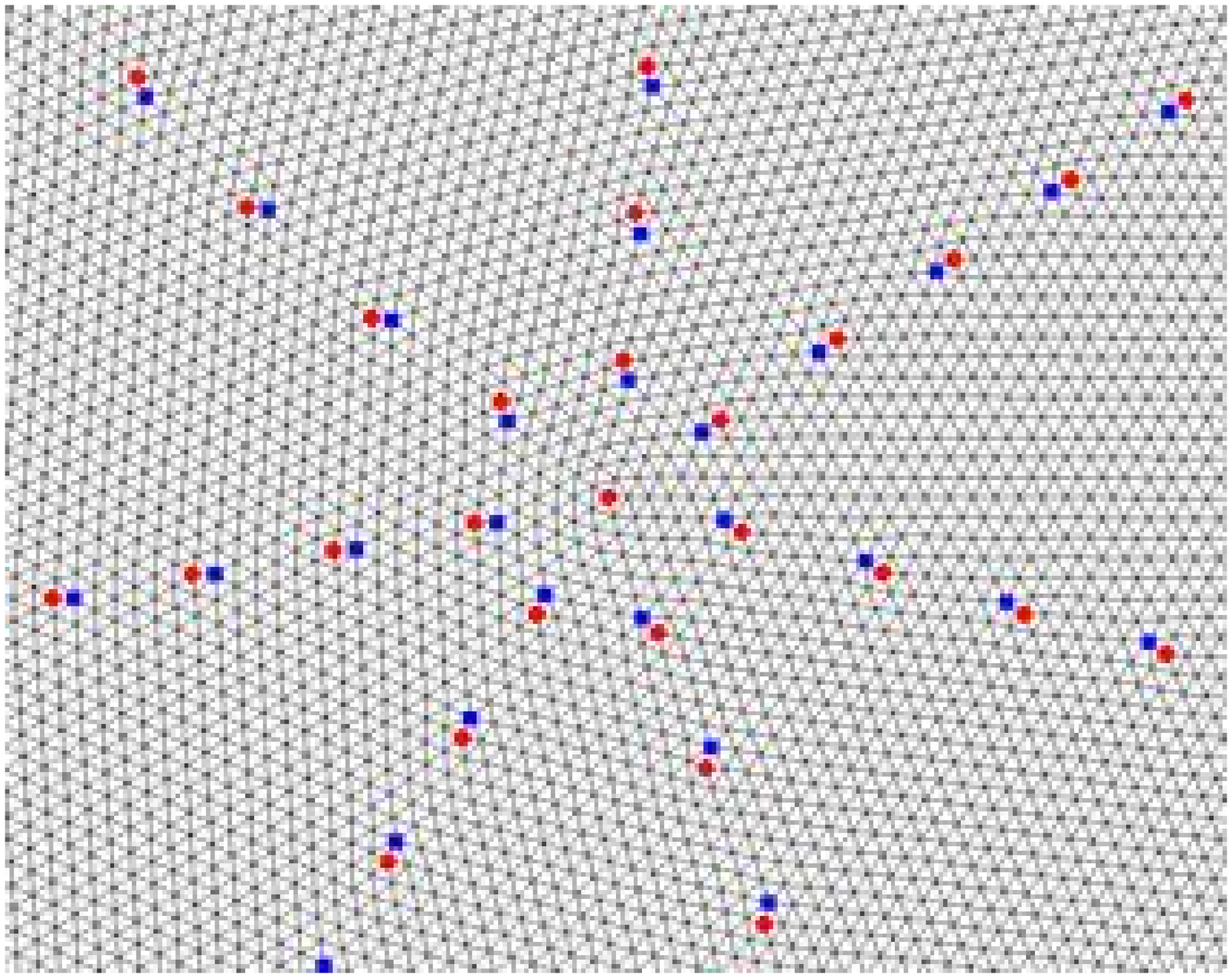}}
\caption{Relaxed configurations for $m=7$, $D=7$ and $\lambda=\mu$.
Results correspond to $R=90$.}
\label{fig__7_7}
\centerline{\includegraphics[width=4in]{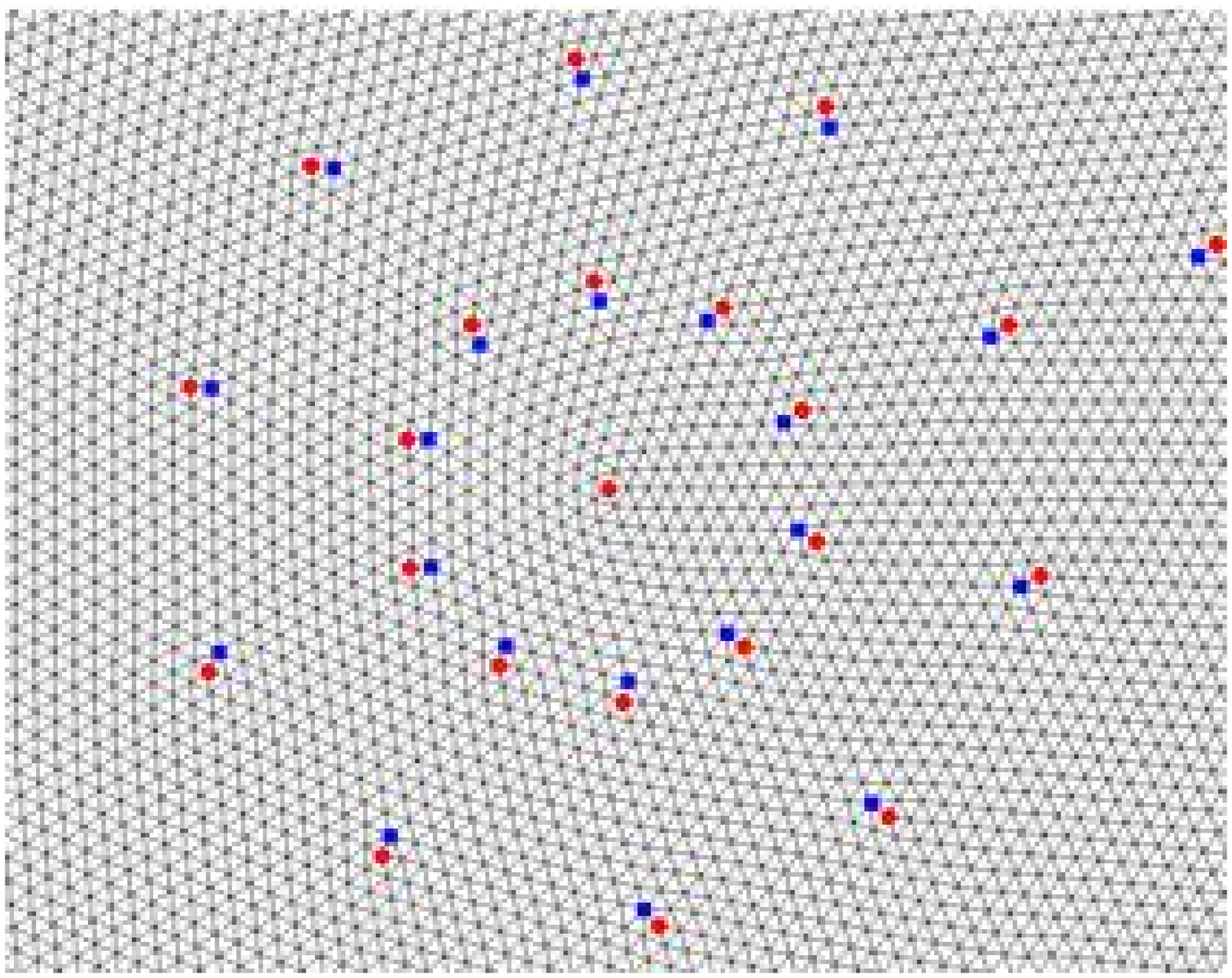}}
\caption{Relaxed configurations for $m=10$, $D=11$ and $\lambda=\mu$.
Results correspond to $R=90$.}
\label{fig__10_11}
\end{figure}

\begin{figure}[ctb]
\centerline{\includegraphics[height=.6\hsize]{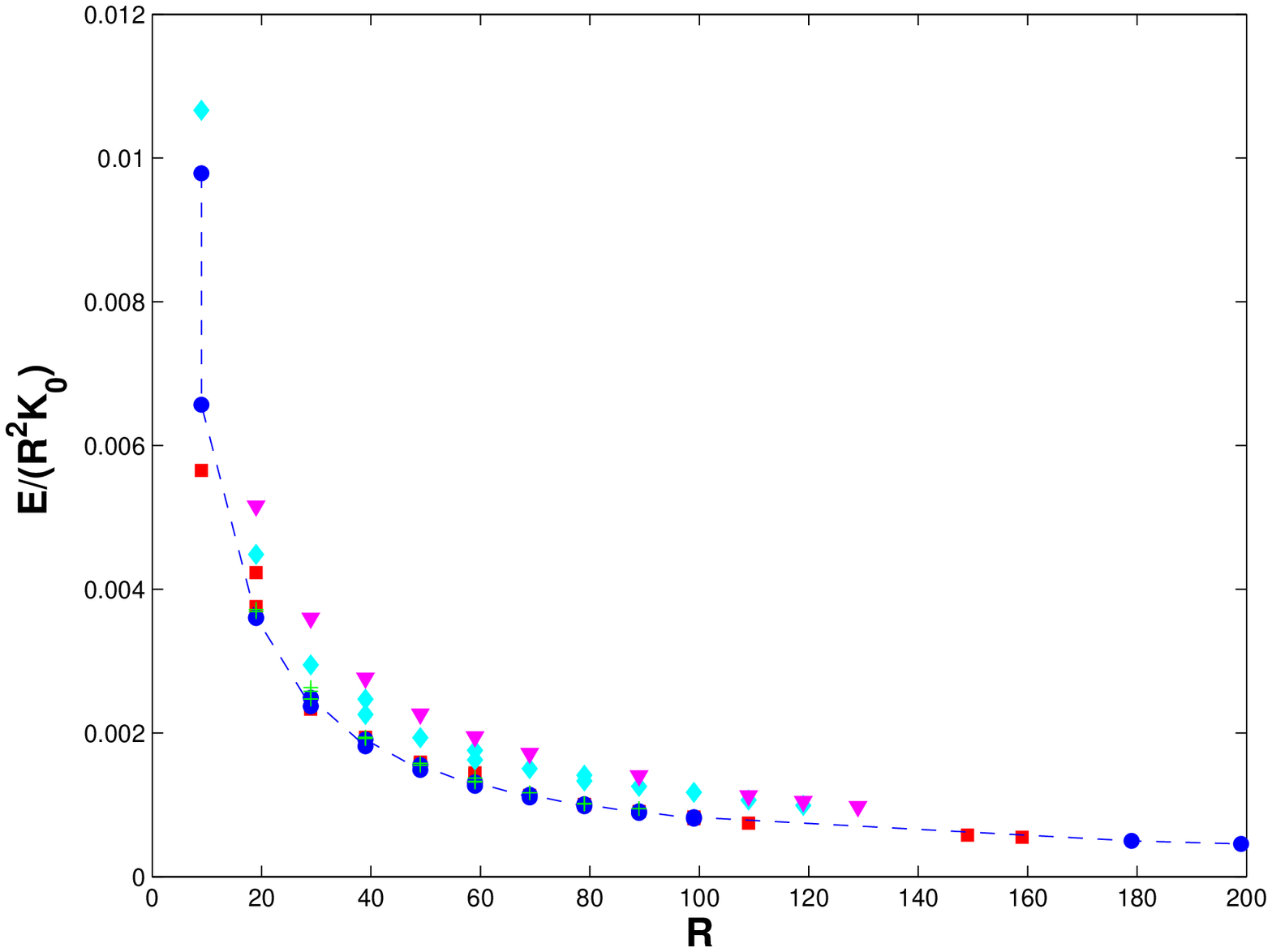}}
\caption{Plot of the energies for $l_{gb} \sim R$ as function of $R$ for $\lambda=\mu$.
$m=3$ (square) $m=4$ (cross) $m=5$ (circle) $m=7$ diamond and $m=10$ (triangle down).}
\label{fig__diff_m}
\end{figure}

The analysis for different value of $m$ will be restricted 
to the optimal spacing values Eq.~\ref{spac__apper} and to the more relevant case
$x \sim 1$ ($l_{gb}\sim R$). Some representative relaxed configurations are
shown in fig.~\ref{fig__3_3}-\ref{fig__10_11}. Similarly as it was found in
the investigation of the dependence on $\Psi$ subsect.~\ref{subsect__or} the constraint
that dislocations can only be oriented along directions defined by the triangular 
lattice is the origin of additional frustration, leading to jagged arrangements which follow the 
ideal orientations only approximately. 

The results corresponding to the energy are shown in fig.~\ref{fig__diff_m}. For finite
radius, the smaller values for $m$ are clearly favored. As $R$ becomes large the values
for the energy become degenerate (within the numerical accuracy) 
in $m$ for $m$ within the range $m=2-5$. An evaluation of the $H$ function 
in Eq.~\ref{arg2} points out that the $m=2$ result should have the lowest energy. However,
the difference in energy for arms $m=2-6$ are small compared with the core energy contribution,
and at this precision, contributions from second order elasticity theory which have not been
included may become important.

For larger $m$ the results show a larger 
energy, which is in qualitative agreement with the logarithmic dependence given
by Eq.~\ref{arg2_largem}. Numerical results for large $m$ become increasingly difficult because
it is difficult to direct the dislocations along the correct directions and also, because
of the small number of dislocations per arm involved. A study including larger volumes 
than the ones performed here is necessary for more rigorous results.
Finite size effects have been predicted to be negligible 
quadratically as a function of $m$ (see Eq.~\ref{arg2}). The convergence is
roughly consistent with that dependence. 

\begin{figure}[hp]
\centerline{\includegraphics[height=.6\hsize]{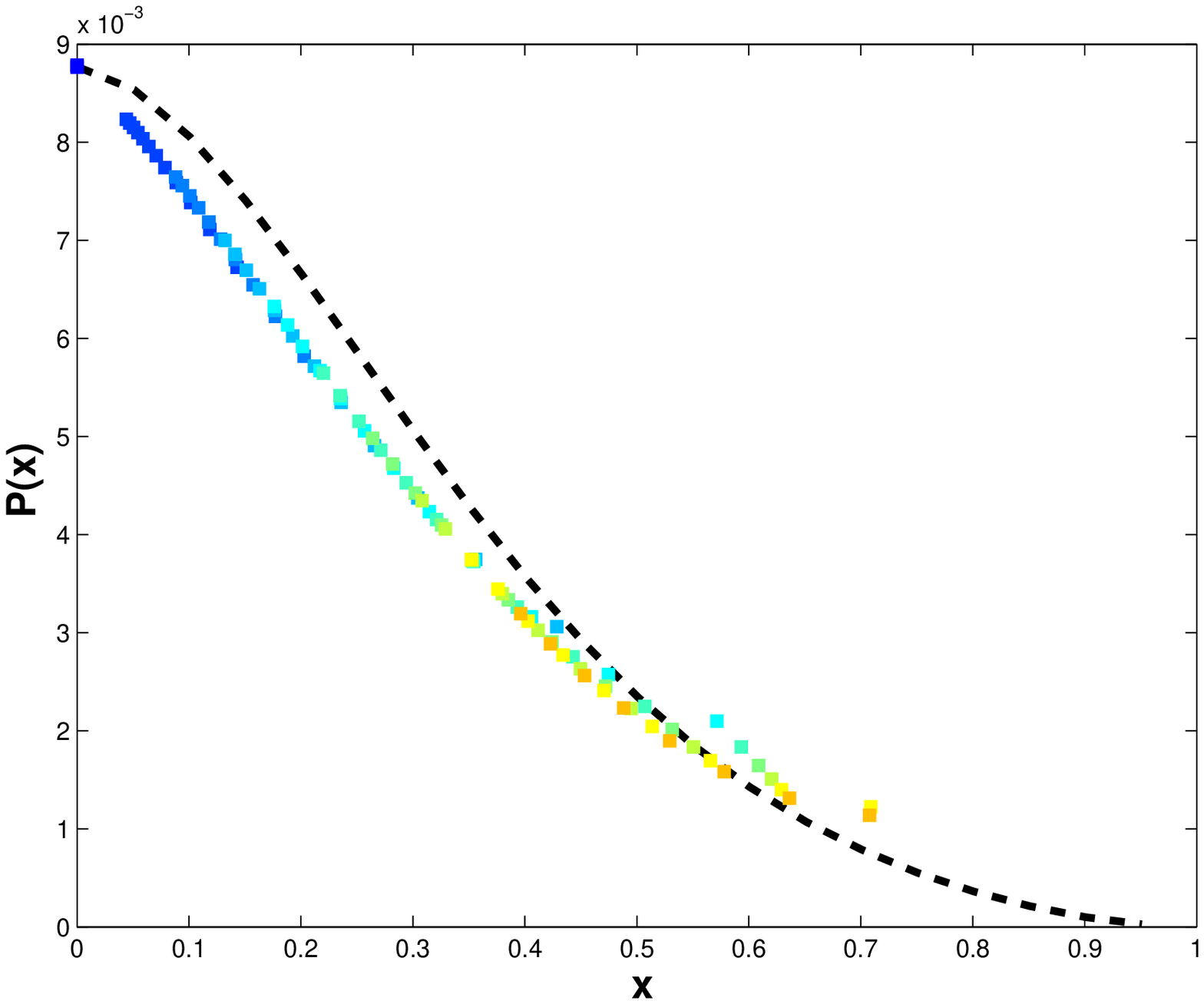}}
\caption{Plot of the scaling function Eq.~\ref{energy_x} for $D=7$, $\lambda=\mu$, 
$m=7$ and $\Psi=\frac{\pi}{7}$ for sizes $R=10-200$. The dashed line is the universal function
for a plus disclination.}  
\label{fig__scaling__spac7__m7}
\centerline{\includegraphics[height=.6\hsize]{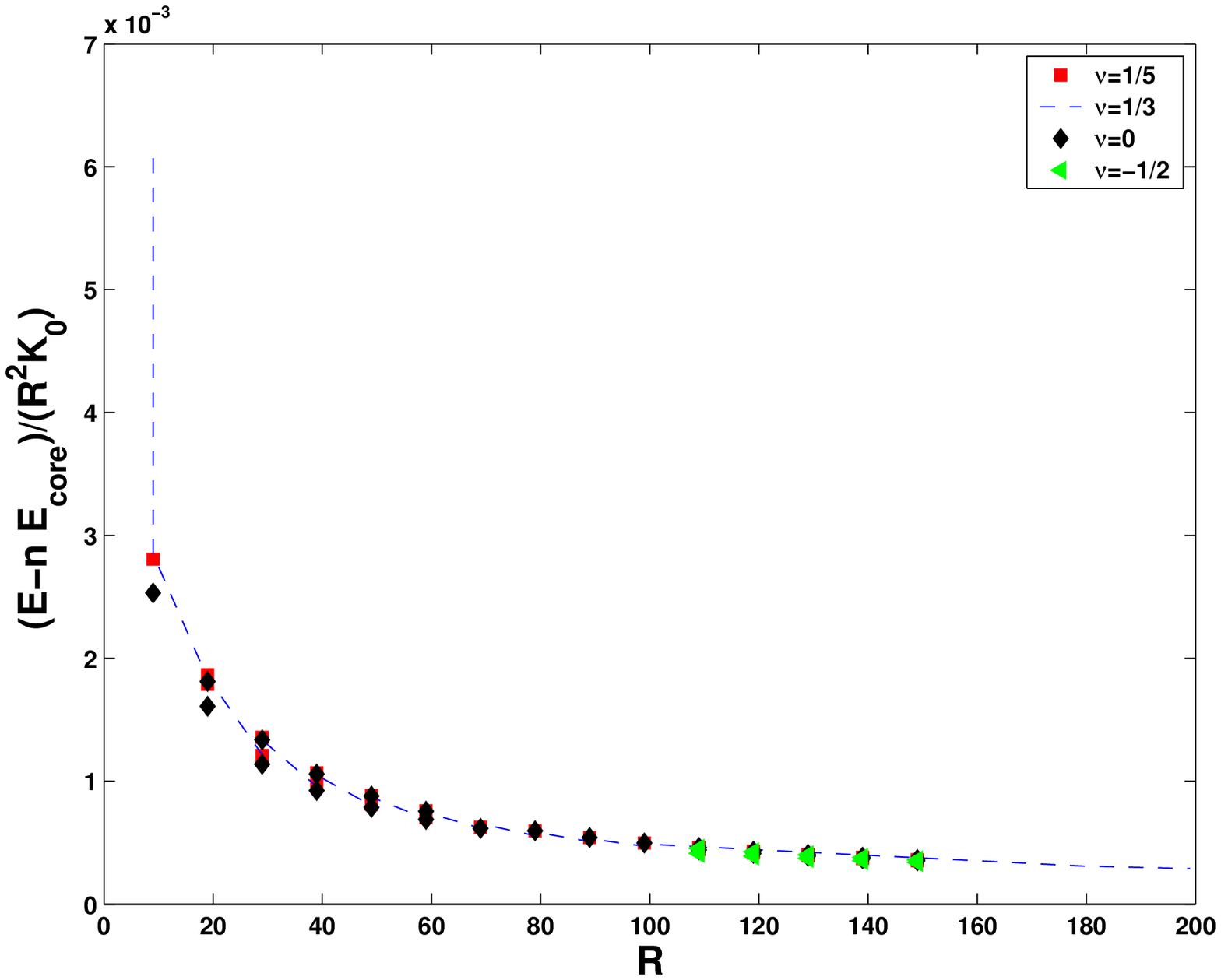}}
\caption{Plot of the energy with the core energy contribution subtracted 
for $l_{gb} \sim R$ as function of $R$ for $m=5$ and
different values of the Poisson ratio $\nu$.}
\label{fig__diff_elastic}
\end{figure}

\subsection{Scaling collapse for minus disclinations}

Results for minus disclinations have not been computed with the same accuracy as for
positive ones. Nevertheless, the same trends as for fivefold defects are
observed, as shown in fig.~\ref{fig__scaling__spac7__m7}. The data 
collapses well to the assumed form Eq.~\ref{energy_x}, and the overall energy
in the intermediate region is smaller than for fivefold defects. 

\section{Universality of the Results}\label{SECT__uni}

The main assumption in the scaling forms for the energy Eq.~\ref{energy_x} 
is that the only dependence on the elastic constants arises from
the Young modulus. This is also true for all sub-leading terms, except for
the core energy coefficient $c$ as it is obvious from table~\ref{Tab__core}. Therefore,
we will present the results by subtracting out the core energy contribution, and the
results should then become universal.

Energy values for $x \sim 1$ plotted this way are shown in
fig.~\ref{fig__diff_elastic}. Except for very small systems,
the universality hypothesis holds very well. Although cannot be conveyed 
from the plot, the energies for different Poisson ratio collapse
to the same universal result with an accuracy less than $0.1\%$. 
This proves that the assumed form for the energy dependence is 
indeed {\em universal}, independent of the microscopic details of the lattice.

This universality of the ${\cal Q}$ function is also remarkable since the present 
calculation goes beyond linear elasticity, not only through 
the quadratic term in displacements kept in the strain tensor 
Eq.~\ref{strain_flat__spac_disc}, but also through the higher powers of the strain
tensor implicitly neglected in going from the discrete energy Eq.~\ref{Disc_Free} 
to the continuum result Eq.~\ref{flat__spac}.
Those non-linear elasticity terms have shown to be quite important when disclinations
are involved, and yet, do not seem to disturb in any way the scaling form 
assumed for the ${\cal Q}$ function.
It seems reasonable then to assume that the results presented are not only {\em universal}
with respect the two elastic constants, but also with respect to higher order terms
in the elastic energy. 

\section{The Spherical Cap}\label{SECT__sph}

\begin{figure}[ctb]
\centerline{\includegraphics[height=.3\hsize]{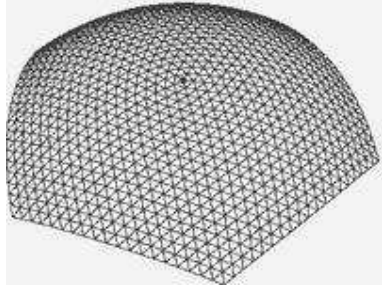}}
\caption{Crystal on an spherical cap. No additional dislocations are needed.}
\label{fig__sphere}
\end{figure}

Although the results presented in this paper are interesting on their own, much of the
motivation for carrying out such a project arises as an effort to provide efficient
computational tools for investigating crystals on frozen geometries. We will 
therefore present a brief outline of some preliminary results for that problem. This will
also provide a broader perspective of the versatility of the method.

Screening of disclination by grain boundaries is not the
only mechanism available if disclinations are allowed to buckle out of the plane:
Gaussian curvature without the need of additional grain boundaries is also a viable
mechanism \cite{SN:88,BowTra:01a}.

An spherical cap of aperture angle $\gamma$ will have a
Gaussian curvature $K=\frac{\gamma^2}{R^2}$. Therefore, for large $\gamma$ the Gaussian
curvature will suffice to screen out the disclination and no additional dislocations will
be needed. The limit of vanishingly small $\gamma$ has been discussed at length in this
paper. Disclinations are screened out by grain boundaries going all the way to the 
boundary of the spherical cap. Therefore, for intermediate values of
$\gamma$, structures of grain boundaries interpolating within these two cases should
be observed \cite{BCNT:02}. 

The result of a minimization for large $\gamma$ is illustrated in fig.~\ref{fig__sphere}.
It is found that, opposite to what happens for the flat case,
additional dislocations actually increase the elastic energy of the system. One should
notice that the triangles next to the boundaries are equilateral, which implies that
the strains are very small (and so is the energy).
A more detailed 
presentation of the results for crystals on spherical caps (positive curvature) including
an investigation of the intermediate regime as well as a similar analysis
for some minimal surfaces (negative curvature) will be presented elsewhere.

\section{Conclusions}\label{SECT__Con}

\subsection{summary of the paper}

The advantages of the computational method presented are:
\begin{enumerate}
\item
Crystals with boundaries can be treated very efficiently.
\item
There are no long range interactions. An entire sweep over the whole system may be
performed with a time proportional to the total volume of the system.
\item
The calculations may be extended to additional geometries with a negligible
additional effort (only introducing the coordinates defining the geometry).
\item
The convergence of the results (at zero temperature) is fast an stable.
\item
The results are very universal, valid for a wide range of potentials. Microscopic details
only enter via the elastic constants.
\end{enumerate}

We provided a very detailed analysis for the problem of the dislocation cloud screening
a disclination. It has been found that the system composed from a
disclination charge $s$ and $m$-radial grain boundaries of dislocations separated a distance
$D$ behaves as a single disclination with an effective charge
\be\label{conc_res}
s_{eff}=s-m\frac{b}{D} \ .
\ee
If $s_{eff}=0$ then the total energy of the system grows linearly with the system
size, similarly as for an infinitely long 
linear grain boundaries composed of dislocations only. The systems 
exhibits a remarkable universality with respect the elastic constants (up to core
energy terms) and higher order elastic terms. The analytical expressions derived 
have been compared with the numerical results, and when disclinations are involved 
the discrepancy occasionally may be as large as $25\%$. Linear elasticity theory does not
provide accurate quantitative estimates for the energetics of some problems
involving disclinations.

\subsection{Implications for Other problems}

We will briefly discuss some of the implications the results found in this paper have
for the problems presented in the introduction. It has been shown that for very small
systems, grain boundaries with the smaller number of arms have the lowest energy. 
This is a general result in agreement with other continuum calculations \cite{BNT:00} in
the context of the sphere. The experimental results in Colloidosomes do show the
same trends \cite{Collo:03}. It would be very interesting to image larger Colloidosomes
and analyze how the defect structures are changed, although that seems difficult in view
of the equilibration times involved.

It has also been shown that if disclinations appear in a crystal phase, 
grain boundaries of dislocations should follow. Isolated disclinations are almost
forbidden at zero temperature. At finite temperatures some properties of the grain 
will change, but for distances close enough to a disclination, the huge
strains that dominate the energy will make additional dislocations inevitable.
Strings ($m=2$ grains) or higher $m$ grains should surround the disclinations,
even at finite temperature. Most of these grains will have a non-zero disclination
charge (positive or negative). This qualitative picture seems in agreement with
recent numerical \cite{SCKCM:97} an experimental results \cite{Tal:89,QG:01}, where the
correlation functions do show agreement with the KTHNY scenario (and therefore with
the unbinding of disclinations) while the typical snapshots of configurations
in equilibrium contain grain boundaries with non-zero disclination charge.
Theories based on grain boundaries \cite{Chui:83} ignore the possibility of having non-zero 
disclination charge from the very beginning. It has been shown in this 
paper that grain boundaries with non-zero disclination charge have energies comparable
as grain boundaries of pure dislocations (with total disclination charge zero), 
provided the spacing within dislocations
is fine tuned appropriately. A complete discussion of the temperature effects is 
obviously of great interest but it is beyond the scope of this paper. 

\subsection{Outlook}

Some issues in this work need either further understanding or
just higher precision data:
\begin{itemize}
\item{Dynamic defect distribution: A global minimization for dislocation positions and
orientations would directly provide the minimum energy configurations. This is a difficult task
since it has been shown that the energy has many almost degenerate local minima.}
\item{Non-integer spacings: Dislocations with non-integer spacings like 
$D=5\frac{\pi}{3}=5.23$, which imply that every third dislocation must be separated an
additional lattice constant, have not been implemented. This may be of some
importance in connection with Eq.~\ref{spac__pred}.}
\item{The degeneracy of the energy as a function of $m$ values should be further refined, since
the accuracy of the results do not allow to discriminate for small values of
$m$.}
\item{The effects of an applied pressure has not been investigated numerically.}
\end{itemize}

The computational method presented can be used to compute the energy of any
distribution of defects in a two dimensional crystal.
Results are in progress to investigate the effects of curvature and
the interactions between disclinations. 

We hope that the considerable detail presented in this paper will not obscure the main results
obtained, but on the contrary, will provide 
convincing arguments for the utility of the present approach.
The code used in this paper will be made publicly available. It may be also requested
from the author.

\bigskip

{\bf ACKNOWLEDGMENTS}

Many parts of this work arised as a result of many clarifying discussions 
with David Nelson, whom I cannot thank enough. I also acknowledge many interesting
discussions with M. Bowick.
I acknowledge S. Rotkin for explaining the interest of this problem 
in the context carbon nanotubes. This work has been supported by Iowa State University
start-up funds.

\appendix
\section{Energy of a disclination}\label{APP_ZERO}

The stresses of a single disclination at the origin may be computed from the Airy function of
a disclination \cite{SN:88}
\be\label{Airy}
\chi=\frac{G s}{2}( A R^2+r^2\log(r)) \ ,
\ee
where $s=\frac{\pi}{3} q_i$,$G=\frac{K_0}{4\pi}$ and $A$ is an undetermined constant.

The stresses for a disclination are given by
\bea\label{stress_discl}
\sigma_{yx}^D&=&-Gs\left[\frac{x y}{r^2}\right]
\nonumber\\
\sigma_{yy}^D&=&Gs\left[\frac{1}{2}+A+\frac{x^2}{r^2}-\log(l_{gb}/r)\right]
\\\nonumber
\sigma_{xx}^D&=&Gs\left[\frac{1}{2}+A+\frac{y^2}{r^2}-\log(l_{gb}/r)\right]  \ ,
\eea
The stress along the radial direction is
\be\label{radial_discl}
\sigma_{r r}=\frac{K_0}{4 \pi}(A+\frac{1}{2}+\log(r))
\ee
therefore, writing $A=-1/2+\frac{4 \pi \Pi}{K_0}-\log(R)$, the stress at the boundary $R$ is
$\sigma_{r r}=\Pi$, and $\Pi$ is interpreted as 
the two dimensional pressure the material.

Plugging Eq.~\ref{stress_discl} into Eq.~\ref{flat__spac}, the energy becomes
\be\label{Sing_Disclin_append}
F=\frac{s^2}{32 \pi}K_0 R^2+\frac{\pi \Pi^2}{8 \cal{B}} R^2,
\ee
where ${\cal B}=\mu+\lambda$ is the two dimensional bulk modulus. 
The result that the stress tensor Eq.~\ref{stress_discl} has an energy
given by the expression Eq.~\ref{Sing_Disclin_append} will be used quite frequently.

\section{Analytical proof of the optimal dislocation spacing}\label{APP_ONE}

We will first compute the stresses generated by a finite grain boundary of length $l_{gb}$,
with dislocations within the grain separated by distance $D$. 

The stresses of a dislocation of Burgers vector $b$ located at the $y$ axis at
point $(0,z)$ are given by
\bea\label{disl_stres}
\sigma_{xy}&=&Gb x \frac{x^2-(y-z)^2}{(x^2+(y-z)^2)^2}
\nonumber\\
\sigma_{yy}&=&-Gb (y-z) \frac{x^2-(y-z)^2}{(x^2+(y-z)^2)^2}
\\\nonumber
\sigma_{xx}&=&Gb (y-z) \frac{3 x^2+(y-z)^2}{(x^2+(y-z)^2)^2} \ ,
\eea
where $G=\frac{K_0}{4\pi}$, where $K_0$ is the Young modulus.
The total stresses will be computed from the methods used in dislocation 
pile-ups \cite{HiLo:91}.  One obtains
\bea\label{pile_up_xx}
\sigma_{xy}&=&\frac{Gb}{D} \int^0_{-l_{gb}} dz  x \frac{x^2-(y-z)^2}{(x^2+(y-z)^2)^2}
\nonumber\\
&=&-\frac{Db}{H}\left[\frac{yx}{y^2+x^2}-\frac{(y+l_{gb})x}{x^2+(y+l_{gb})^2}\right]
\eea
A similar calculation yields
\bea\label{pile_up_xy}
\sigma_{yy}&=&-\frac{Gb}{D}\left[-\frac{x^2}{y^2+x^2}+\frac{x^2}{x^2+(y+l_{gb})^2}
+\frac{1}{2}\log(\frac{x^2+(y+l_{gb})^2}{x^2+y^2})\right] 
\\\nonumber
\sigma_{xx}&=&-\frac{Gb}{D}\left[\frac{x^2}{y^2+x^2}-\frac{x^2}{x^2+(y+l_{gb})^2}
+\frac{1}{2}\log(\frac{x^2+(y+l_{gb})^2}{x^2+y^2})\right]. 
\eea
If the grain is now rotated and angle $\theta$ the stresses at distances $r<< l_{gb}$ will 
be (where $r=\sqrt(x^2+y^2$):
\bea\label{pile_up_rotate}
\sigma_{yx}&=&-\frac{Gb}{D}\left[\sin\theta\cos\theta+\frac{x y}{r^2}\right]
\nonumber\\
\sigma_{yy}&=&-\frac{Gb}{D}\left[-\frac{x^2\cos^2\theta}{r^2}+
\frac{y^2\sin^2\theta}{r^2}+\log(l_{gb}/r)\right]
\\\nonumber
\sigma_{xx}&=&-\frac{Gb}{D}\left[\frac{x^2\cos^2\theta}{r^2}-\frac{y^2\sin^2\theta}{r^2}
+\log(l_{gb}/r)\right]
\eea
Adding together $m$ arms separated an angle $\frac{2\pi}{m}$, one obtains
\bea\label{pile_up_total}
\sigma_{yx}^G&=&-m\frac{Gb}{D}\left[\frac{x y}{r^2}\right]
\nonumber\\
\sigma_{yy}^G&=&-m\frac{Gb}{D}\left[\frac{-x^2+y^2}{2r^2}+\log(l_{gb}/r)\right]
\\\nonumber
\sigma_{xx}^G&=&-m\frac{Gb}{D}\left[\frac{x^2-y^2}{2r^2}+\log(l_{gb}/r)\right]  \ ,
\eea
where the identity 
$\sum_{l=1}^{m}\cos^2(\frac{2\pi l}{m})=\sum_{l=1}^{m}\sin^2(\frac{2\pi l}{m})=m/2$ has been
used.

The stresses for a disclination have already been computed \ref{stress_discl} with result
\bea\label{pile_up_discl}
\sigma_{yx}^D&=&-Gs\left[\frac{x y}{r^2}\right]
\nonumber\\
\sigma_{yy}^D&=&Gs\left[\frac{1}{2}+A+\frac{x^2}{r^2}-\log(l_{gb}/r)\right]
\\\nonumber
\sigma_{xx}^D&=&Gs\left[\frac{1}{2}+A+\frac{y^2}{r^2}-\log(l_{gb}/r)\right]  \ .
\eea
The total stresses are
\bea\label{pile_up_sum}
\sigma_{yx}^D&=&-G(s-\frac{b}{D}m)\left[\frac{x y}{r^2}\right]
\nonumber\\
\sigma_{yy}^D&=&G(s-\frac{b}{D}m)\left[\frac{x^2}{r^2}-\log(l_{gb}/r)\right]+\Pi
\\\nonumber
\sigma_{xx}^D&=&G(s-\frac{b}{D}m)\left[\frac{y^2}{r^2}-\log(l_{gb}/r)\right]+\Pi  \ ,
\eea
where the external pressure $\Pi$ is given by $\Pi=((A-1/2)s+\frac{b}{2 H}m) G$. 
Using Eq.~\ref{Sing_Disclin_append} for the modified disclination charge 
$s_{eff}=s-\frac{b}{D}m$, one finds
\be\label{new_energy}
F=\frac{s_{eff}^2}{32 \pi}K_0 R^2+ \frac{\pi \Pi^2}{8 \pi {\cal B}} R^2 ,
\ee
Therefore, perfect screening implies $s_{eff}=0$, 
\be\label{condit_der}
s=\frac{b}{D}m
\ee
and there is no external pressure $\Pi=0$, the system of disclination plus grain boundary does
diverge quadratically with the system size. It can also be proved that within the same
approach, the term linear in system size vanishes also.
It should be emphasized that Eq.~\ref{condit_der} is a linear order result.
One should expect that if all orders were included, some modifications should occur, 
among them formula Eq.~\ref{condit_der} should become Eq.~\ref{spac__pred} 

\section{Energy of a Grain boundary of dislocations}\label{APP__GB}

The energy of a linear grain boundary of $n$ dislocations with Burgers vector
perpendicular to the grain is a well known result \cite{RS:50}. Here we just 
outline the aspects important for this paper. The total energy is given by
\be\label{tot_ener_d_gb}
E=\frac{K_0 a^2}{8\pi} n \int d^2 {\bf r} \sigma_{xy} \ ,
\ee
where $\sigma_{xy}$ is the stress tensor of the dislocations in the $xy$ direction. 
The final result of the energy is
\be\label{fin_en_d_gb}
E= \frac{K_0 a^2}{4\pi D} H(\frac{2 \pi a}{D}) R \ ,
\ee
where $a$ is the lattice constant and
\be\label{H_d_gb}
H(x)=-\log(1-e^{-x})+\frac{x}{e^x-1} \ ,
\ee
for small $x$, (dislocations separated a distance $D >> a$), the usual formula
$H(x)=-\log(x)+1$ follows. If $x$ is not small enough, the formula
Eq~\ref{H_d_gb} must be used.

\section{Relations of base vectors}

If ${\bf e}_1$ and ${\bf e}_2$ are the triangular defining the triangular 
lattice the following relations follow.
\be\label{app_double}
\sum_{b} e^{i}_{ab} e^{j}_{ab} = 3 \delta^{i j} \ ,
\ee
where the sum is over all nearest neighbors $b$. It also follows
\be\label{app_four_a}
\sum_{b} e^{i}_{ab} e^{j}_{ab} e^{k}_{ab} e^{l}_{ab}=
\frac{6}{8}(\delta^{ij}\delta^{kl}+\delta^{li}\delta^{kj}+
\delta^{ki}\delta^{lj})
\ee
and
\be\label{app_four_b}
\sum_{b} e^{i}_{ab} e^{j}_{ab} e^{k}_{a b+1} e^{l}_{a b+1}=
\frac{6}{8}(\delta^{ij}\delta^{kl}+R^{li}R^{kj}+
R^{lj}R^{ki}) \ ,
\ee
where $R^{ij}=\cos(\pi/3)\delta^{ij}+\sin(\pi/3)\epsilon^{ij}$. Similarly,
\be\label{app_four_c}
\sum_{b} e^{i}_{ab} e^{j}_{a b+1} e^{k}_{ab} e^{l}_{a b+1}=
\frac{6}{8}(R^{ji} R^{lk}+R^{li}R^{jk}+
\delta^{ki}\delta^{lj}) \ ,
\ee
and
\be\label{app_four_d}
\sum_{b} e^{i}_{ab} e^{j}_{a b} e^{k}_{ab} e^{l}_{a b+1}=
\frac{6}{8}(\delta^{ji} R^{lk}+R^{li}\delta^{jk}+
\delta^{ik}R^{lj}) \ .
\ee

\bibliography{/home/trvsst/Research/Bibliography/Grain,/home/trvsst/Research/Bibliography/General,/home/trvsst/Research/Bibliography/reviews,/home/trvsst/Research/Bibliography/melt,/home/trvsst/Research/Bibliography/Langmuir}

\end{document}